\documentclass[twocolumn,amsmath,amssymb,pra,superscriptaddress]{revtex4-1}
\usepackage[dvipdfmx]{graphicx}
\usepackage{dcolumn}
\usepackage{bm}
\usepackage{subfigure}
\usepackage{booktabs}
\usepackage{color}
\newcounter{one}
\def\bra#1{\mbox{\boldmath $#1$}^{\top}}
\def\ket#1{\mbox{\boldmath $#1$}}
\newcommand{\bracket}[1]{\left\langle #1 \right\rangle}

\newcommand{\affA}{
Artificial Intelligence Research Center, 
National Institute of Advanced Industrial Science and Technology, 
2-3-26 Aomi, Koto-ku, Tokyo, Japan
}

\begin{document}

\title{\textbf{Algorithmic detectability threshold of the stochastic block model}}
\author{Tatsuro Kawamoto}
\affiliation{\affA}

\begin{abstract}
The assumption that the values of model parameters are known or correctly learned, i.e., the Nishimori condition, is one of the requirements for the detectability analysis of the stochastic block model in statistical inference. In practice, however, there is no example demonstrating that we can know the model parameters beforehand, and there is no guarantee that the model parameters can be learned accurately. In this study, we consider the expectation--maximization (EM) algorithm with belief propagation (BP) and derive its algorithmic detectability threshold. Our analysis is not restricted to the community structure, but includes general modular structures. Because the algorithm cannot always learn the planted model parameters correctly, the algorithmic detectability threshold is qualitatively different from the one with the Nishimori condition. 
\end{abstract}
\maketitle

\section{Introduction}
Graph clustering is a technique to detect a macroscopic law of connectivity in a graph \cite{Goldenberg2010,Leger2013,Peixoto2017tutorial,Fortunato2010,Fortunato2016}. 
In other words, we expect that each vertex in a graph belongs to a module (or modules) and that the vertices in the same module are statistically equivalent. We then let an algorithm infer the most likely module assignments. 
In particular, the subset of the graph-clustering problem that focuses on the detection of densely connected (i.e., assortative) module is termed community detection. 
A classic example of community detection is the detection of social groups in a social network, wherein the vertices and edges represent persons and friendships, respectively. 

Often, the graph structures that are more general than the assortative modules are detected using statistical inference methods. 
In this approach, we infer the most-likely module assignment for each vertex by fitting graph data using a random graph model with a planted modular structure. 
The stochastic block model (SBM) \cite{holland1983stochastic,FienbergMeyerWasserman85,WangWong87,AndersonWassermanFaust92,FaustWasserman92,KarrerNewman2011} is a canonical model used for this purpose. 
The SBM is an extension of the Erd\H{o}s-R\'{e}nyi random graph; each vertex in the SBM has a planted module assignment and the vertices in the same module have stochastically equivalent connection patterns. 
Thus, we can generate the graph instances of various modular structures. 

The model parameters of the SBM smoothly connect the random graphs with a strong modular structure and the Erd\H{o}s-R\'{e}nyi random graph. 
Interestingly, as the strength of the modular structure decreases, before the SBM becomes equivalent to the Erd\H{o}s-R\'{e}nyi random graph, a phase transition occurs, and at this stage, it becomes impossible to infer the planted module assignment. 
This critical point is called the \textit{detectability threshold} or the \textit{detectability limit} \cite{Reichardt2008,Decelle2011,NussinovPRE2012,banks2016information,Young2017}. 
The impossibility of inference stems from the fact that the fluctuations of the graph instances are not negligible, so that the graph instances generated from the SBM are statistically indistinguishable from the those of the Erd\H{o}s-R\'{e}nyi random graph. 
This is a fundamental problem in graph clustering, and it offers an insight into the extent to which we should expect algorithms to work. 
This is a characteristic phenomenon of sparse graphs, i.e., graphs with a constant average degree, and it cannot be observed in dense graphs. 
In the dense regime, instead, another interesting problem called the recovery problem \cite{Condon2001,BickelChenPNAS2009,Rohe2011,yun2014community,AbbeNIPS2015,AbbleReview2017} arises. 

\begin{figure}[t]
 \begin{center}
   \includegraphics[width=\columnwidth]{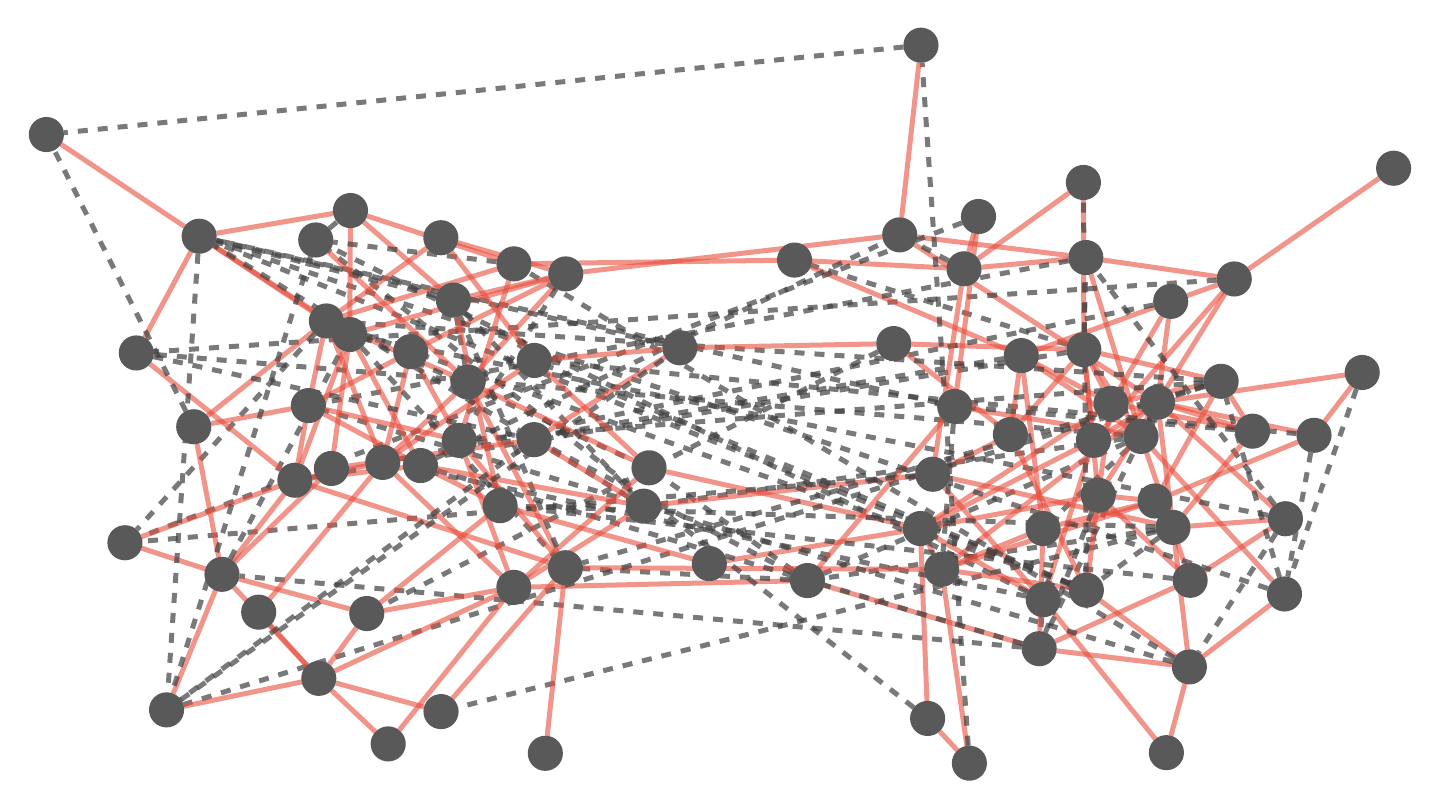}
 \end{center}
 \caption{
	(Color online)  An instance of the labeled SBM with two modules. 
	The red solid edges represent the edges with an assortative structure, and the gray dashed edges represent the edges with a disassortative structure. 
	}
 \label{labeledSBMinstance}
\end{figure}

Throughout this paper, we focus on sparse undirected graphs in the infinite size limit. 
We do not consider graphs with self-loops and multi-edges or the SBM wherein a vertex belongs to multiple modules. 
Instead, we allow the graphs to have multiple types of edges \cite{Heimlicher2012,Lelarge2015,Mariadassou2010,GuimeraPloS13,RoviraSciRep13,Saade2016,PeixotoPRE2017}. 
Thus, we will extend our analysis to a variant of the SBM called the labeled stochastic block model (labeled SBM). 
An instance of the labeled SBM is shown in Fig.~\ref{labeledSBMinstance}. 

Several frameworks corresponding to algorithms such as the greedy \cite{KarrerNewman2011,Larremore2014}, Monte Carlo \cite{Nowicki2001,PeixotoPRE2014MonteCarlo,PeixotoPRX2014,NewmanReinert2016}, and expectation--maximization (EM) \cite{daudin08,latouche12,Ball2011,Decelle2011a,Aicher2014,NewmanClauset2016} algorithm are available for the statistical inference of the SBM. 
We analyze the performance of the EM algorithm, because it is scalable and suited for theoretical analysis. 
In particular, we consider belief propagation (BP) as its module-assignment inference (E-step), and the point estimate of the model parameters as its model-parameter learning (M-step). 

In this study, we derive the algorithmic detectability threshold of the SBM using the EM algorithm. 
The detectability threshold is often defined as the fundamental limit where all polynomial-time algorithms fail; here, we distinguish such a threshold as the theoretical limit of detectability. 
There are two reasons why we focus on the algorithmic detectability threshold. 

The first reason is that it helps deriving a legitimate threshold in practice. 
It is known that the Bayesian inference using BP achieves the theoretical limit of detectability \cite{Decelle2011,Decelle2011a,Mossel2015,Massoulie2014,MooreReview2017}, assuming that the correct values of model parameters are known. 
Supported by this favorable property in theory and the good scalability of the algorithm, BP was implemented to solve various SBM variants \cite{Aicher2014,ZhangPRE2014,Ghasemian2016,NewmanClauset2016,KawamotoGMGM2017}. 
On the other hand, some nonBayesian methods are known to be strictly suboptimal, e.g., \cite{KawamotoKabashimaPRE2015,KawamotoKabashimaEPL2015,Javanmard2016}. 
However, this is not a fair comparison because the Bayesian inference with BP is not an algorithmic detectability threshold; it can be regarded as the EM algorithm without the M-step. 
In practice, we do not know the planted model parameters beforehand, and there is no guarantee that they can be learned precisely. 
In fact, the performance of the EM algorithm generally depends on the initial values of the model parameters. 
To the best of our knowledge, the algorithmic detectability threshold that takes into account both the E-step and M-step remains a mystery, and we solve this problem analytically. 
The condition that the planted values of model parameters are correctly learned is often referred to as the Nishimori condition \cite{Iba1999,Decelle2011a,LenkaFlorentReview2016}. 

The second reason why we focus on the algorithmic detectability threshold is that we need to observe the algorithmic infeasibility through the analysis. 
In general, even if we have data of extremely high dimensions, there exists a limit that an algorithm can handle correctly. 
Then, the inference task is algorithmically infeasible though it may be computationally feasible. 
This difficulty can be depicted in the detectability phase-diagram of the algorithm, whereas it cannot be observed from the theoretical limit of detectability because of the assumption of infinite learnability. 

This paper is organized as follows. 
In Sec.~\ref{Sec:SBM}, we explain the precise construction of the standard (i.e., binary label) SBM and its analytically tractable parametrization. 
We then explain the procedure of the EM algorithm (Sec.~\ref{Sec:EMalgorithm}) and the behavior of its M-step (Sec.~\ref{Sec:MstepTransientDynamics}). 
From Sec.~\ref{Sec:labeledSBM}, we extend the SBM to the labeled SBM. 
After explaining how the formulation of the standard SBM is modified in the labeled SBM in Sec.~\ref{Sec:labeledSBM}, we derive the algorithmic detectability threshold in Sec.~\ref{Sec:AlgorithmicDetectabilityThreshold}. 
In Sec.~\ref{Sec:DetectabilityPhaseDiagram}, we present the detectability phase-diagrams for some specific cases. 
In Sec.~\ref{Sec:AlgoInfeasibility}, we discuss the algorithmic infeasibility as a physical consequence of the algorithmic detectability threshold. 
Finally, Sec.~\ref{Sec:SummaryDiscussion} is devoted to the summary and discussion. 
While we focused on a limited case of the same problem in Ref.~\cite{ADT-Lett}, in this study, we extend the results therein as general as possible.

\section{Standard SBM}\label{Sec:SBM}
We first explain how the standard SBM is generated. 
We consider the set of vertices $V$ with $|V| = N$ and denote the number of modules as $q$. 
For each vertex, we assign a module label $\sigma \in \{1,\dots,q\}$ with probability $p(\sigma \lvert \ket{\gamma}) = \gamma_{\sigma}$ independently and randomly, where $\ket{\gamma}$ is an array that determines the relative size of each module. 
We denote an array of module assignments as $\ket{\sigma}$. 
Given $\ket{\sigma}$, an undirected edge is generated between vertices $i$ and $j$ independently with probability $p(A_{ij}=1 \lvert \sigma_{i},\sigma_{j}, \ket{c}) = c_{\sigma_{i} \sigma_{j}}/N$, where $A_{ij}$ is the adjacency matrix element with $A_{ij} = 1$ if $i$ and $j$ are connected, and $A_{ij} = 0$ if they are not connected (or connected via a nonedge). 
The matrix $\ket{c}$ is called the affinity matrix. 
It is a $q \times q$ matrix and is of $O(1)$ so that the resulting graph is sparse. 
We denote an edge between vertices $i$ and $j$ as $(i,j)$, the set of edges as $E$, and the number of edges as $L$. 
Thus, the likelihood of the standard SBM is as follows. 
\begin{align}
p(A, \ket{\sigma} \lvert \ket{\gamma},\ket{c}) 
&= p(\ket{\sigma} \lvert \ket{\gamma}) p(A \lvert \ket{\sigma}, \ket{c}) \notag\\
&= \prod_{i=1}^{N} \gamma_{\sigma_{i}} \prod_{i<j} \left( \frac{c_{\sigma_{i}\sigma_{j}}}{N} \right)^{\delta_{A_{ij},1}} \left(1 - \frac{c_{\sigma_{i}\sigma_{j}}}{N} \right)^{\delta_{A_{ij},0}}. \label{SBMlikelihood}
\end{align}
The model parameters to be learned in the SBM are $\ket{\gamma}$ and $\ket{c}$, and the set of module assignments $\ket{\sigma}$ is the latent variable that is to be inferred, given the adjacency matrix $A$.

Although the SBM is very flexible, it is sometimes difficult to treat it analytically. 
Therefore, it is common to restrict the affinity matrix to the simple community structure $c_{\sigma\sigma^{\prime}} = c_{\mathrm{in}}$ for $\sigma = \sigma^{\prime}$, and $c_{\sigma\sigma^{\prime}} = c_{\mathrm{out}}$ otherwise. 
However, this parametrization largely restricts the graph ensemble that the SBM can originally express. 
Therefore, we instead consider the following affinity matrix \cite{KawamotoGMGM2017,Young2017}. 
\begin{align}
\ket{c} &= \Delta c \, W + c_{\mathrm{out}} \ket{1}\bra{1}, \label{affinityGMGM}
\end{align}
where $\Delta c \equiv c_{\mathrm{in}} - c_{\mathrm{out}}$, $\ket{1}$ is a column vector with all elements equal to unity. $W$ is an indicator matrix in which $W_{\sigma \sigma^{\prime}} = 1$ represents the densely connected module pair, and $W_{\sigma \sigma^{\prime}} = 0$ otherwise; the simple community structure is the special case where $W$ is the identity matrix. 
While this model is parametrized only by $c_{\mathrm{in}}$ and $c_{\mathrm{out}}$, it can express various modular structures by the choices of $W$, which is the input. 
We focus on undirected graphs, and thus, $W$ is symmetric. 

As the SBM has the average degree $c = \bra{\gamma} \ket{c \gamma}$, the affinity matrix of Eq.~(\ref{affinityGMGM}) can also be parametrized by $c$ and $\Delta c$. 
Note that because both $c_{\mathrm{in}}$ and $c_{\mathrm{out}}$ are nonnegative, $\Delta c/c$ is bounded as 
\begin{align}
-\frac{1}{1 - \Omega} \le \frac{\Delta c}{c} \le \frac{1}{\Omega}, 
\end{align}
where we defined $\Omega \in (0,1)$ as $\Omega \equiv \bra{\gamma} W \ket{\gamma}$. 
In the following paragraphs, we consider the average degree $c$ as the input and parametrize the strength of the modular structure by a normalized parameter $x \in [0,1]$ that linearly interpolates the maximum and minimum values of $\Delta c/c$: 
\begin{align}
x &= \Omega \left[ 1 + (1-\Omega)\frac{\Delta c}{c} \right]. \label{xDefinition}
\end{align}
The graph exhibits assortative and disassortative structures when $x > \Omega$ and $x < \Omega$, respectively.

\section{Statistical inference of the SBM}\label{Sec:EMalgorithm}
In principle, the model parameters $\ket{\gamma}$ and $\ket{c}$ are learned by maximizing the marginalized log-likelihood $p(A \lvert \ket{\gamma}, \ket{c})=\sum_{\ket{\sigma}}p(A, \ket{\sigma} \lvert \ket{\gamma},\ket{c})$ and the set of module assignments $\ket{\sigma}$ is determined by the posterior distribution $p(\ket{\sigma} \lvert A, \ket{\gamma}, \ket{c})$. 
However, because their exact computation is demanding, we use the EM algorithm for approximation. 

Using the variational expression, $\log p(A \lvert \ket{\gamma}, \ket{c})$ can be expressed as 
\begin{align}
\log p(A \lvert \ket{\gamma}, \ket{c}) 
&= \mathbb{E}_{\psi}\left[\log \frac{p(A, \ket{\sigma} \lvert \ket{\gamma}, \ket{c})}{\psi(\ket{\sigma})}\right] \notag\\
&\hspace{50pt}+ D_{\mathrm{KL}}\left( \psi(\ket{\sigma}) || p(\ket{\sigma} \lvert A, \ket{\gamma}, \ket{c})\right), \label{VariationalExpression}
\end{align}
where $\psi(\ket{\sigma})$ is our estimate of the module-assignment distribution, and $\mathbb{E}_{\psi}[\cdots]$ is the corresponding average. 
The second term $D_{\mathrm{KL}}(P || Q)$ represents the Kullback--Leibler divergence of distributions $P$ and $Q$. 
Equation (\ref{VariationalExpression}) indicates that if our estimate $\psi(\ket{\sigma})$ coincides with the posterior distribution $p(\ket{\sigma} \lvert A, \ket{\gamma}, \ket{c})$, then the model parameters can be learned by maximizing the first term of Eq.~(\ref{VariationalExpression}). 
Note that this is a double-optimization problem because $p(\ket{\sigma} \lvert A, \ket{c})$ is conditioned on $\ket{c}$. 
Therefore, the EM algorithm iteratively updates the module-assignment inference (E-step) and the model-parameter learning (M-step). 
We use the hat notation for the estimated (learned) value of the model parameter. 
Note also that we do not explicitly need the whole joint distribution of $\ket{\sigma}$. 
Because the SBM only has pair-wise interactions, as confirmed from the calculation of the M-step, we only need to calculate the one-point and two-point marginals of the posterior estimates.

Although the E-step can be performed in various ways, we employ the BP algorithm. 
It solves for the marginal distributions of module assignments based on the tree approximation \cite{Yedidia_NIPS2000,Yedidia2003,MezardMontanari2009,Decelle2011a}. 
This approximation is justified because we consider the sparse graphs, which are locally tree-like. 
We estimate the marginal distribution $\psi^{i}_{\sigma}$ of the module assignment of vertex $i$ as follows. 
\begin{align}
\psi^{i}_{\sigma} 
&= \frac{\hat{\gamma}_{\sigma}}{Z^{i}} 
\prod_{\ell \notin \partial i } 
\left( 1 - \sum_{\sigma_{\ell}} \psi^{\ell \to i}_{\sigma_{\ell}} \frac{\hat{c}_{\sigma_{\ell}\sigma}}{N} \right)
\prod_{k \in \partial i} 
\left( \sum_{\sigma_{k}} \psi^{k \to i}_{\sigma_{k}} \hat{c}_{\sigma_{k}\sigma} \right) \notag\\
&\simeq \frac{\hat{\gamma}_{\sigma}}{Z^{i}} \exp\left[-\sum_{\ell=1}^{N} \sum_{\sigma_{\ell}} \psi^{\ell}_{\sigma_{\ell}} \frac{\hat{c}_{\sigma_{\ell}\sigma}}{N} \right] \prod_{k \in \partial i} \left( \sum_{\sigma_{k}} \psi^{k \to i}_{\sigma_{k}} \hat{c}_{\sigma_{k}\sigma} \right), \label{BPSBM1}
\end{align}
where $\partial i$ represents the neighboring vertices of $i$; $Z^{i}$ is the normalization factor; and we use the sparse approximation. 
In the factor of neighboring vertices, $\psi^{k \to i}_{\sigma}$ represents the marginalized distribution of vertex $k$ with missing knowledge of edge $(i,k)$. 
Although the elements $\{\psi^{k \to i}_{\sigma}\}$ in Eq.~(\ref{BPSBM1}) are correlated to each other, in general, we can treat them independently when the graph is tree-like. 
Analogously to Eq.~(\ref{BPSBM1}), $\psi^{i \to j}_{\sigma}$ is calculated as 
\begin{align}
\psi^{i \to j}_{\sigma} 
&\simeq \frac{\hat{\gamma}_{\sigma}}{Z^{i \to j}} \exp\left[-\sum_{\ell=1}^{N} \sum_{\sigma_{\ell}} \psi^{\ell}_{\sigma_{\ell}} \frac{\hat{c}_{\sigma_{\ell}\sigma}}{N} \right] \notag\\
&\hspace{60pt} \times\prod_{k \in \partial i \backslash j} \left( \sum_{\sigma_{k}} \psi^{k \to i}_{\sigma_{k}} \hat{c}_{\sigma_{k}\sigma} \right), \label{BPSBM2}
\end{align}
where $k \in \partial i \backslash j$ is the set of neighbors of $i$ where vertex $j$ is excluded, and $Z^{i \to j}$ is a normalization factor. 
Note that Eq.~(\ref{BPSBM2}) constitutes of a set of closed equations, and so we can iteratively update the values of $\{\psi^{i \to j}_{\sigma}\}$. 
The BP algorithm updates Eq.~(\ref{BPSBM2}) until convergence, and it calculates the complete marginal $\psi^{i}_{\sigma}$ by Eq.~(\ref{BPSBM1}).

We next explain the M-step. 
Because we consider the affinity matrix $\ket{c}$ that is parametrized by $c$ and $\Delta c$, the only parameter that we need to update is $\Delta c$ and $\ket{\gamma}$. 
From Eqs.~(\ref{SBMlikelihood}) and (\ref{affinityGMGM}), the extremum point of the first term of Eq.~(\ref{VariationalExpression}) can be calculated analytically. 
Following Ref.~\cite{KawamotoGMGM2017}, we obtain 
\begin{align}
\Delta\hat{c} = \frac{N^{2}}{2} \frac{N \sum_{(i,j) \in E} \mathbb{E}_{\psi}\left[W_{\sigma_{i} \sigma_{j}}\right] - c \sum_{i<j}\mathbb{E}_{\psi}\left[W_{\sigma_{i} \sigma_{j}}\right]}{\sum_{i<j}\mathbb{E}_{\psi}\left[W_{\sigma_{i} \sigma_{j}}\right] \sum_{i<j}\left(1 - \mathbb{E}_{\psi}\left[W_{\sigma_{i} \sigma_{j}}\right]\right)}. \label{DeltacUpdate1}
\end{align}
Here we have 
\begin{align}
\mathbb{E}_{\psi}\left[W_{\sigma_{i} \sigma_{j}}\right] 
&= \sum_{\sigma_{i} \sigma_{j}} W_{\sigma_{i} \sigma_{j}} \frac{\psi^{i \to j}_{\sigma_{i}} \hat{c}_{\sigma_{i} \sigma_{j}} \psi^{j \to i}_{\sigma_{j}}}{\sum_{\sigma_{i} \sigma_{j}} \psi^{i \to j}_{\sigma_{i}} \hat{c}_{\sigma_{i} \sigma_{j}} \psi^{j \to i}_{\sigma_{j}}} \notag\\
&= \frac{(c + \Delta \hat{c} (1 - \hat{\Omega})) X^{ij}}{c + \Delta \hat{c}(X^{ij} - \hat{\Omega})}, \label{meanW}
\end{align}
where we defined $X^{ij} \equiv \ket{\psi}^{i \to j} W \ket{\psi}^{j \to i \top}$; $\ket{\psi}^{i \to j}$ is the row vector $(\psi^{i \to j}_{1}, \dots, \psi^{i \to j}_{q})$.
As often assumed \cite{Decelle2011a,ZhangMartinNewman2015,KawamotoGMGM2017}, if there is no macroscopic fluctuation with respect to the number of vertices in each module, we can approximate that 
\begin{align}
\frac{2}{N^{2}} \sum_{i<j} \mathbb{E}_{\psi}\left[W_{\sigma_{i} \sigma_{j}}\right] \simeq \bra{\hat{\gamma}} W \ket{\hat{\gamma}} = \hat{\Omega}. \label{MacroApprox}
\end{align}
Using Eqs.~(\ref{meanW}) and (\ref{MacroApprox}), Eq.~(\ref{DeltacUpdate1}) is rewritten as 
\begin{align}
\Delta \hat{c}^{(t+1)} = \frac{c}{\hat{\Omega} (1 - \hat{\Omega})} 
\left[ \bracket{\frac{\left( c + \Delta \hat{c}^{(t)} \left( 1 - \hat{\Omega} \right) \right) X^{ij}}{c + \Delta \hat{c}^{(t)} \left(X^{ij} - \hat{\Omega}\right)}}_{E} - 1 \right], \label{DeltacUpdate2}
\end{align}
where $\bracket{Y_{ij}}_{E} \equiv L^{-1} \sum_{(i,j) \in E} Y_{ij}$. 
Here, we introduced superscript $(t)$ to indicate the $t$th update. 

For estimating each element in $\ket{\gamma}$, the extremum condition of the first term of Eq.~(\ref{VariationalExpression}) readily yields 
\begin{align}
\hat{\gamma}_{\sigma} = \frac{1}{N} \sum_{i=1}^{N} \psi^{i}_{\sigma}. 
\end{align}

\section{Transient dynamics of the M-step}\label{Sec:MstepTransientDynamics}
\begin{figure}[t]
 \begin{center}
   \includegraphics[width=\columnwidth]{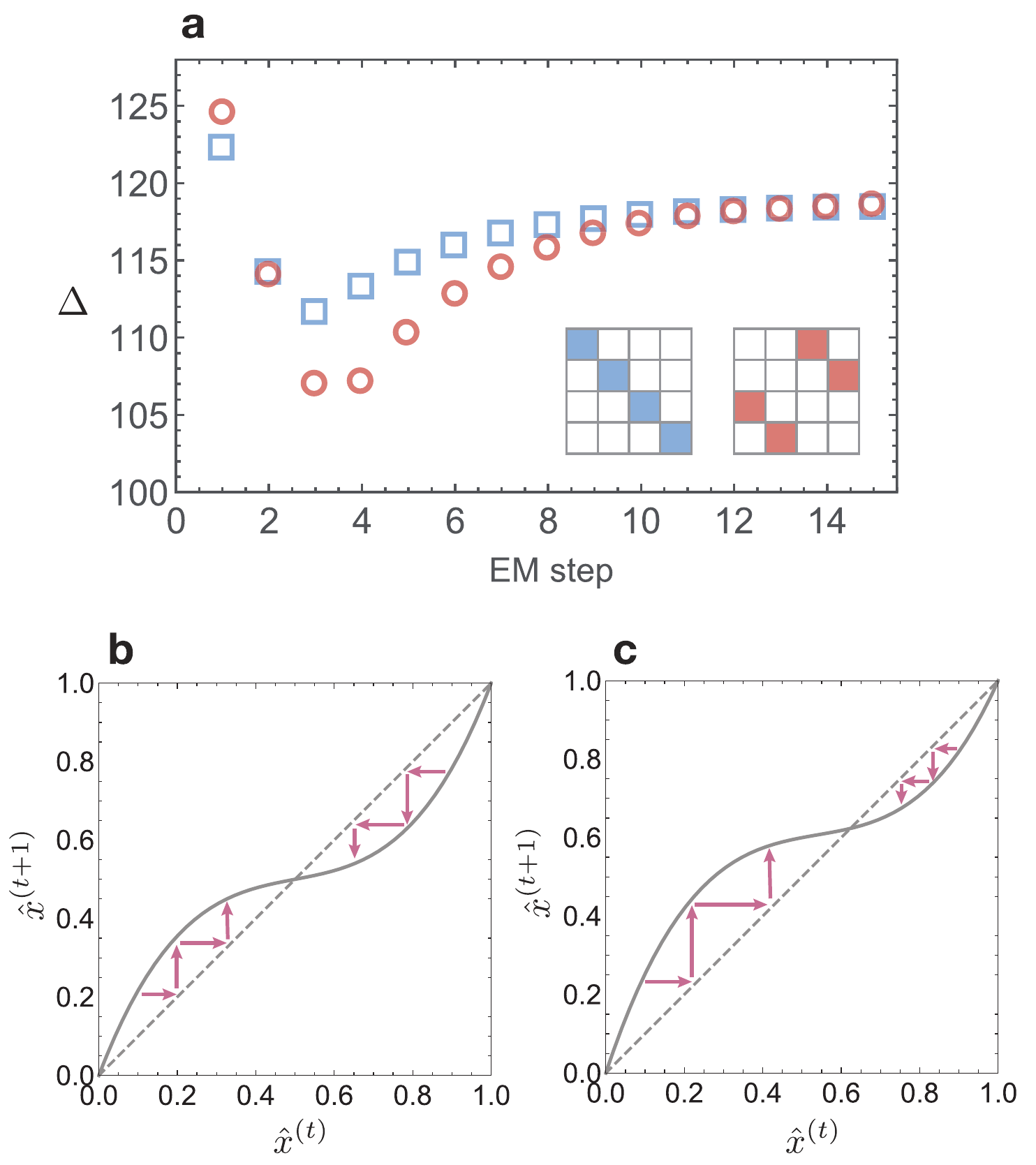}
 \end{center}
 \caption{
	(Color online) (top) (a) Learning curves of the model parameters that indicate the strength of the modular structures. 
	The vertical axis represents the total variation $\Delta$ of the affinity matrix elements $c_{\sigma \sigma^{\prime}}$ from the mean value $\overline{\ket{c}}$. 
	The horizontal axis represents the number of steps in the EM algorithm. 
	We considered examples for the SBMs with a simple community structure (circles) and a bipartite structure (rectangles), as shown in the inset. 
	To detect these structures, we did not use the affinity matrix restricted to Eq.~(\ref{affinityGMGM}); instead, we used the one with full degrees of freedom. 
	(bottom) Second-order expansions of the right-hand side of Eq.~(\ref{xUpdate2}) with the first moments ({\bf b}) $\bracket{\xi_{ij}} = 0$ and ({\bf c}) $\bracket{\xi_{ij}} = 0.2$ (gray curves). 
	For the other parameters, we set $\Omega = 0.5$ and $\bracket{\xi^{2}_{ij}} = 0.9$. 
	The dashed line represents $\hat{x}^{(t+1)} = \hat{x}^{(t)}$. 
	The arrows in each figure show the update process of $\hat{x}^{(t)}$ schematically. 
	}
 \label{SBMtransient}
\end{figure}

The transient dynamics of the M-step is the key to deriving the algorithmic detectability threshold. 
The trajectories of the model parameter updates for the standard SBM are exemplified in Fig.~\ref{SBMtransient}a. 
The vertical axis represents the total variation $\Delta \equiv \sum_{\sigma, \sigma^{\prime}} |c_{\sigma \sigma^{\prime}} - \overline{\ket{c}}|$ from the mean value $\overline{\ket{c}} \equiv \sum_{\sigma, \sigma^{\prime}} c_{\sigma \sigma^{\prime}}/q^{2}$; $\Delta = 0$ indicates the uniform structure. 
An important observation here is that the model parameter estimate is not attracted directly to the planted value. 
Instead, they are attracted to the point of uniform structure first. 
The EM algorithm encounters the algorithmic detectability threshold during this transient regime. 

To gain a deeper insight about the nonlinear update equation (\ref{DeltacUpdate2}), we express it in terms of $\hat{x}$. 
\begin{align}
\hat{x}^{(t+1)} &= \hat{x}^{(t)} \bracket{\frac{X^{ij}}{\hat{\Omega} + \frac{\hat{x}^{(t)}-\hat{\Omega}}{1-\hat{\Omega}}(X^{ij} - \hat{\Omega}) } }_{E}. \label{xUpdate1}
\end{align}
Note that when $\ket{\psi}^{i \to j} = \ket{\hat{\gamma}}$ for any $(i,j)$, i.e., when BP does not provide any additional knowledge compared to the prior distribution, we have $X^{ij} = \hat{\Omega}$, so $\hat{x}$ will not be updated. 
Here, we introduce the normalized deviation $\xi_{ij} \equiv (X^{ij} - \hat{\Omega})/\hat{\Omega}$ and rewrite Eq.~(\ref{xUpdate1}) as 
\begin{align}
\hat{x}^{(t+1)} 
&= \hat{x}^{(t)} \bracket{ \frac{1+\xi_{ij}}{1 + \frac{\hat{x}^{(t)}-\hat{\Omega}}{1-\hat{\Omega}}\xi_{ij} } }_{E}. \label{xUpdate2}
\end{align}
Because we usually have no prior information about the distribution of the marginals, it is common to set $\ket{\psi}^{i \to j}$ uniformly random; i.e., $\bracket{\xi_{ij}}_{E} = 0$ and $\bracket{\xi^{2}_{ij}}_{E} > 0$ at the beginning of the algorithm. 
(In Appendix~\ref{PsiTransientDynamics}, we show that this condition may be relaxed for the case of equally sized modules.)

Because the estimate of the module sizes $\hat{\ket{\gamma}}$ are also updated concurrently, the M-step can be very complicated in general. 
Fortunately, however, the update dynamics for $\hat{\ket{\gamma}}$ can be neglected for the analysis in this study. 
When we derive the detectability threshold in Sec.~\ref{Sec:AlgorithmicDetectabilityThreshold}, we need to restrict ourselves to the case of equal-size modules. 
In that case, as shown in Appendix~\ref{PsiTransientDynamics}, the distribution $\ket{\psi}^{i \to j}$ is kept randomized during the transient regime, and it yields the estimate that the module sizes are equal.

The specific shape of Eq.~(\ref{xUpdate2}) is shown in Fig.~\ref{SBMtransient}b. 
The fixed points $\hat{x}^{(t+1)} = \hat{x}^{(t)}$ are located at $x=0, 1$, and $x=\hat{\Omega}$. 
The former two fixed points indicate the parameters of the bipartite graph and the graph with completely disconnected modules, respectively. 
The last fixed point indicates the parameter of the uniform random graph. 
From Fig.~\ref{SBMtransient}b, we can confirm that $x=0$ and $x=1$ are unstable fixed points, while $x=\hat{\Omega}$ is a stable fixed point. 
Note that Eq.~(\ref{xUpdate2}) is independent of the input graph during the transient regime. 
Moreover, because Eq.~(\ref{xUpdate2}) corresponds to an arbitrary modular structure, this tendency holds irrespective of the specific structure that we assume in the model. 
Therefore, the M-step of the EM algorithm exhibits universal dynamics at the beginning of the algorithm. 
When the transient regime is over, the dynamics is no longer universal, and $\hat{x}$ moves toward the planted value as long as it is detectable [Fig.~\ref{SBMtransient}c].

\section{Labeled SBM}\label{Sec:labeledSBM}
Before we consider the algorithmic detectability threshold, we extend the standard SBM to the labeled SBM that has $p$ types of edges, i.e., $A_{ij} = \alpha \in \{0, \dots, p\}$; $\alpha=0$ represents the nonedge. 
We denote the set of $\alpha$-edges as $E_{\alpha}$, and therefore, $E = \cup_{\alpha=1}^{p} E_{\alpha}$ ($|E_{\alpha}| = L_{\alpha}$, $L = \sum^{p}_{\alpha=1} L_{\alpha}$). 
Analogously to Eq.~(\ref{SBMlikelihood}), the likelihood of the labeled SBM can be expressed as 
\begin{align}
p(A, \ket{\sigma} \lvert \ket{\gamma}, \ket{c}^{\alpha}) 
&= \prod_{i=1}^{N} \gamma_{\sigma_{i}} \prod_{i<j} \prod_{\alpha = 0}^{p} \left( \frac{c^{\alpha}_{\sigma_{i}\sigma_{j}}}{N} \right)^{\delta_{A_{ij},\alpha}}, \label{labeledSBMlikelihood}
\end{align}
where the affinity matrix element $c^{\alpha}_{\sigma_{i}\sigma_{j}}$ with respect to $\alpha$-edges obeys the normalization constraint $N^{-1} \sum_{\alpha=0}^{p} c^{\alpha}_{\sigma \sigma^{\prime}} = 1$ for any $\sigma$ and $\sigma^{\prime}$. 
We denote the average degree and the strength of the modular structure of $\alpha$-edges as $c_{\alpha}$ and $\Delta c_{\alpha}$, respectively ($\sum_{\alpha>0}c_{\alpha} = c$). 
We also denote the fraction of $\alpha$-edges as $P_{\alpha}$, i.e., $P_{\alpha} \equiv c_{\alpha}/c$. 

Corresponding to the likelihood Eq.~(\ref{labeledSBMlikelihood}), we can extend the BP update equation (\ref{BPSBM2}) to the one for the labeled SBM in a straightforward manner, as follows. 
\begin{align}
\psi^{i \to j}_{\sigma} 
&= \frac{\gamma_{\sigma}}{Z^{i \to j}} 
\prod_{\ell \in \left[\partial i \backslash j\right]^{0}} 
\left( 1 - \sum_{\alpha>0} \sum_{\sigma_{\ell}} \psi^{\ell \to i}_{\sigma_{\ell}} \frac{\hat{c}^{\alpha}_{\sigma_{\ell}\sigma}}{N} \right) \notag\\
&\hspace{60pt} \times\prod_{\alpha>0} \prod_{k \in \left[\partial i \backslash j\right]^{\alpha}} 
\left( \sum_{\sigma_{k}} \psi^{k \to i}_{\sigma_{k}} \hat{c}^{\alpha}_{\sigma_{k}\sigma} \right) \notag\\
&\simeq \frac{\gamma_{\sigma}}{Z^{i \to j}} 
\prod_{\alpha>0} \varphi^{i \to j}_{\alpha, \sigma}, \label{BPlabeledSBM1}
\end{align}
where $\varphi^{i \to j}_{\alpha, \sigma}$ is the $\alpha$-edge generalization of Eq.~(\ref{BPSBM2}), 
\begin{align}
\varphi^{i \to j}_{\alpha, \sigma} 
&= \exp\left[-\sum_{\ell} \sum_{\sigma_{\ell}} \psi^{\ell}_{\sigma_{\ell}} \frac{\hat{c}^{\alpha}_{\sigma_{\ell}\sigma}}{N} \right] \prod_{k \in \left[\partial i \backslash j\right]^{\alpha}} \left( \sum_{\sigma_{k}} \psi^{k \to i}_{\sigma_{k}} \hat{c}^{\alpha}_{\sigma_{k}\sigma} \right). \label{BPlabeledSBM2}
\end{align}
The vertex $k \in \left[\partial i \backslash j\right]^{\alpha}$ is a neighbor of $i$ such that $(i,k) \in E_{\alpha}$ and $k \ne j$.

It is also straightforward to generalize Eq.~(\ref{xUpdate2}) to the labeled SBM. 
Because the variables of different edge labels are not directly coupled, we can treat them separately. 
We can define $x_{\alpha}$ analogously to $x$ in Eq.~(\ref{xDefinition}), and for each label $\alpha$, 
\begin{align}
\hat{x}^{(t+1)}_{\alpha} 
&= \hat{x}^{(t)}_{\alpha} \bracket{ \frac{1+\xi^{\alpha}_{ij}}{1 + \frac{\hat{x}^{(t)}_{\alpha}-\hat{\Omega}_{\alpha}}{1-\hat{\Omega}_{\alpha}}\xi^{\alpha}_{ij} } }_{\alpha}, \label{xUpdate3}
\end{align}
where $\bracket{Y_{ij}}_{\alpha} \equiv L_{\alpha}^{-1} \sum_{(i,j) \in E_{\alpha}} Y_{ij}$. 
Corresponding to $W^{\alpha}$ for each $\alpha$, $\hat{\Omega}$ and $X^{ij}$ are generalized to $\hat{\Omega}_{\alpha} \equiv \bra{\hat{\gamma}} W^{\alpha} \ket{\hat{\gamma}}$ and $X^{ij}_{\alpha} \equiv \ket{\psi}^{i \to j} W^{\alpha} \ket{\psi}^{j \to i \top}$. Accordingly, we defined $\xi^{\alpha}_{ij} \equiv (X^{ij}_{\alpha} - \hat{\Omega}_{\alpha})/\hat{\Omega}_{\alpha}$.

\section{Detectability threshold}\label{Sec:AlgorithmicDetectabilityThreshold}
We now derive the algorithmic detectability threshold of the EM algorithm. 
As in other inference-based detectability analyses, here, we need to impose further restrictions. 
We hereafter focus on the case where the module size is equal, i.e., $\gamma_{\sigma} = 1/q$ for any $\sigma$, and the average degree of each module is equal, i.e., $\sum_{\sigma^{\prime}}W_{\sigma\sigma^{\prime}} = a$ ($a = \text{const.}$) for any $\sigma$. 
In addition, we assume that the $W$ matrix is common for all $\alpha$. 
In other words, while the edges of different types may indicate assortative and disassortative structures, they share the same planted modules. 

The undetectable phase can be characterized as the phase in which we cannot retrieve any information about the planted module assignments through the BP update equation (\ref{BPlabeledSBM1}). 
In the present setting, it is equivalent to the condition where the \textit{factorized state} is a stable fixed point of BP \cite{Decelle2011a,MezardMontanari2009,MooreReview2017}. 
The factorized state has the form $\ket{\psi}^{i \to j}$ such that $\psi^{i \to j}_{\sigma} = 1/q$ for any $\sigma$ and $(i,j)$, i.e., the state exhibits no signal of a likely-module assignment for any vertex.

\subsection{Instability of the factorized state}\label{InstabilityOfFactorizedState}
The instability condition of the factorized state in a sparse graph, which is often termed the Kesten--Stigum bound, can be analyzed using the framework of tree reconstruction \cite{JansonMossel2004,MezardMontanari2009,Decelle2011a,MooreReview2017}. 
We assume that the graph is a tree and evaluate whether perturbations from the leaves are significant or negligible for the inference of the root vertex. 
We denote $v_{0}$ as the root vertex and $v_{i}$ as the descendant vertices in $i$th generation. 
When the vertices at distance $d$ are perturbed as $\ket{\epsilon}_{d}$, the variation $\delta \psi^{v_{0}}_{\sigma_{v_{0}}}$ of the marginal probability at the root vertex $v_{0}$ is expressed as follows. 
\begin{widetext}
\begin{align}
\delta \psi^{v_{0}}_{\sigma_{v_{0}}} \left( \ket{\epsilon}_{d} \right) 
&= \sum_{v_{1} \in \partial v_{0}} \sum_{\sigma_{v_{1}}} \frac{\delta \psi^{v_{0}}_{\sigma_{v_{0}}}}{\delta \psi^{v_{1} \to v_{0}}_{\sigma_{v_{1}}}} \delta \psi^{v_{1} \to v_{0}}_{\sigma_{v_{1}}} \notag\\
&= \sum_{v_{1} \in \partial v_{0}} \sum_{\sigma_{v_{1}}} \frac{\delta \psi^{v_{0}}_{\sigma_{v_{0}}}}{\delta \psi^{v_{1} \to v_{0}}_{\sigma_{v_{1}}}} \sum_{v_{2} \in \partial v_{1}\backslash v_{0}} \sum_{\sigma_{v_{2}}} \frac{\delta \psi^{v_{1} \to v_{0}}_{\sigma_{v_{1}}}}{\delta \psi^{v_{2} \to v_{1}}_{\sigma_{v_{2}}}} \delta \psi^{v_{2} \to v_{1}}_{\sigma_{v_{2}}}\notag\\
&= \sum_{(v_{1}, v^{\prime}_{1})\in E} \sum_{\sigma_{v_{1}}} \delta_{v^{\prime}_{1}v_{0}} \frac{\delta \psi^{v_{0}}_{\sigma_{v_{0}}}}{\delta \psi^{v_{1} \to v^{\prime}_{1}}_{\sigma_{v_{1}}}} 
\sum_{(v_{2}, v^{\prime}_{2})\in E} \sum_{\sigma_{v_{2}}} \delta_{v^{\prime}_{2}v_{1}}(1 - \delta_{v^{\prime}_{1}v_{2}}) \frac{\delta \psi^{v_{1}\to v^{\prime}_{1}}_{\sigma_{v_{1}}}}{\delta \psi^{v_{2} \to v^{\prime}_{2}}_{\sigma_{v_{2}}}} \delta \psi^{v_{2} \to v^{\prime}_{2}}_{\sigma_{v_{2}}}\notag\\
&= \sum_{(v_{1}, v^{\prime}_{1})\in E} \sum_{\sigma_{v_{1}}} \delta_{v^{\prime}_{1}v_{0}} \frac{\delta \psi^{v_{0}}_{\sigma_{v_{0}}}}{\delta \psi^{v_{1} \to v^{\prime}_{1}}_{\sigma_{v_{1}}}} 
\sum_{(v_{2}, v^{\prime}_{2})\in E} \sum_{\sigma_{v_{2}}} B_{v_{1}\to v^{\prime}_{1}, v_{2}\to v^{\prime}_{2}} T^{v_{1}\to v^{\prime}_{1}, v_{2}\to v^{\prime}_{2}}_{\sigma_{v_{1}} \sigma_{v_{2}}} \notag\\
&\hspace{20pt} \dots \sum_{(v_{d}, v^{\prime}_{d})\in E} \sum_{\sigma_{v_{d}}} B_{v_{d-1}\to v^{\prime}_{d-1}, v_{d}\to v^{\prime}_{d}} T^{v_{d-1}\to v^{\prime}_{d-1}, v_{d}\to v^{\prime}_{d}}_{\sigma_{v_{d-1}} \sigma_{v_{d}}} \epsilon^{v_{d}\to v^{\prime}_{d}}_{\sigma_{v_{d}}}, \label{stability-1}
\end{align}
\end{widetext}
where we defined the nonbacktracking matrix $B$ \cite{Krzakala2013} and the transfer matrix $T$ as 
\begin{align}
& B_{i\to i^{\prime}, j\to j^{\prime}} = \delta_{i j^{\prime}}(1 - \delta_{i^{\prime} j}) \hspace{20pt} (i\to i^{\prime}, j\to j^{\prime} \in E), \\
& T^{i\to i^{\prime}, j\to j^{\prime}}_{\sigma_{i} \sigma_{j}} = \frac{\delta \psi^{i\to i^{\prime}}_{\sigma_{i}}}{\delta \psi^{j \to j^{\prime}}_{\sigma_{j}}}. 
\end{align}
In general, Eq.~(\ref{stability-1}) cannot be expressed as the tensor product of matrices $B$ and $T$ because $T$ depends on the edge label. 
However, because we assume that all the affinity matrices $\ket{c}^{\alpha}$ share the common $W$ matrix, the transfer matrix $T$ for each edge type differs only by a constant factor. 
Then, we can express Eq.~(\ref{stability-1}) as 
\begin{align}
\delta \psi^{v_{0}}_{\sigma_{v_{0}}} \left( \ket{\epsilon}_{d} \right) 
&\simeq \left[ \left( B^{\prime} \otimes T^{\prime} \right)^{d} \ket{\epsilon}_{d} \right]_{(v_{0}\to v^{\prime}_{0}), \sigma_{v_{0}}}. \label{stability-2}
\end{align}
In the case of a simple community structure, when the perturbation from the factorized state is considered \cite{Decelle2011a,Krzakala2013,ZhangMoore2014}, the elements of $B^{\prime}$ and $T^{\prime}$ are 
\begin{align}
B^{\prime}_{i\to i^{\prime}, j\to j^{\prime}} &= \frac{\Delta \hat{c}_{\alpha}}{q c_{\alpha}} B_{i\to i^{\prime}, j\to j^{\prime}} \hspace{10pt} \left( \alpha = A_{j j^{\prime}} \right), \label{Bprime}\\
T^{\prime}_{\sigma_{i}\sigma_{j}} &= \delta_{\sigma_{i}\sigma_{j}} - \frac{1}{q}.  \label{Tprime}
\end{align}
For inferring the general modular structure, we analyze the BP algorithm of the transformed basis $\ket{\Psi}^{i \to j} \equiv \ket{\psi}^{i \to j}W$ as considered in Ref.~\cite{KawamotoGMGM2017}. 
In this case, the transfer matrix $T^{\prime}$ is given by 
\begin{align}
T^{\prime}_{\sigma_{i}\sigma_{j}} &= W_{\sigma_{i}\sigma_{j}} - \Omega.  \label{TprimeGMGM}
\end{align}

From Eq.~(\ref{stability-2}), the instability condition attributed to the perturbation is determined by the eigenvalues of $T^{\prime}$ and $B^{\prime}$. 
Note that because the unit vector $\ket{1}$ is a leading eigenvector of $W$, we have $|\lambda_{1}(T^{\prime})| = |\lambda_{2}(W)|$, where $\lambda_{1}(T^{\prime})$ is the leading eigenvalue of $T^{\prime}$ and $\lambda_{2}(W)$ is the second-leading eigenvalue of $W$, which may be degenerated with the leading eigenvalue.
Therefore, unless all the elements in the leading eigenvector of $B^{\prime}$ have the same sign, the instability condition of the factorized state, i.e., the detectable region is determined by 
\begin{align}
&|\lambda_{2}(W)| \left| \lambda_{1}\left( B^{\prime} \right) \right| > 1.
\end{align}
When all the elements in the leading eigenvector of $B^{\prime}$ do have the same sign, e.g., the sign of $\Delta c_{\alpha}$ is the same for all $\alpha$, the leading eigenvector is unrelated to the modular structure. In this case, the second-leading eigenvalue determines the detectable region, i.e., 
\begin{align}
&|\lambda_{2}(W)| \left| \lambda_{2}\left( B^{\prime} \right) \right| > 1. 
\end{align}

The eigenvalue of $B^{\prime}$ that we should refer to is determined by the boundary of the spectral band, which we denote as $\lambda_{\mathrm{b}}$, and the isolated eigenvalue, which we denote as $\lambda_{\mathrm{iso}}$. 
Note that the eigenvalues of $B^{\prime}$ changes dynamically because it is a function of the estimated value of the model parameters. 
When we initially assume a strong modular structure, i.e., large values of $|\Delta \hat{c}_{\alpha}|$, $|\lambda_{2}(W)| |\lambda_{\mathrm{b}}| > 1$ holds. 
Then, the spectral band shrinks because of the universal dynamics of the M-step, until $|\lambda_{2}(W)| |\lambda_{\mathrm{b}}|$ reaches unity. (For a specific example, see Sec.~\ref{TwoCommunityPhaseDiagram} and Fig.~\ref{EMtrajectories-q2}.)
At this stage, the factorized state stabilizes if there is no isolated eigenvalue that is correlated to the planted modular structure. 
As we mentioned earlier, $\hat{x}_{\alpha}$ will no longer be updated once the factorized state is achieved. 
Hence, $|\lambda_{2}(W)| |\lambda_{\mathrm{b}}| = 1$ determines the values of $\Delta\hat{c}_{\alpha}$ that we should refer to at the detectability threshold. 
Given this estimate, the factorized state becomes unstable when $|\lambda_{2}(W)| |\lambda_{\mathrm{iso}}| > 1$ is satisfied. 
Hence, the boundary condition of the detectable phase is given by 
\begin{align}
& |\lambda_{2}(W)| |\lambda_{\mathrm{b}}| = 1, 
\hspace{20pt} |\lambda_{2}(W)| |\lambda_{\mathrm{iso}}| = 1. \label{ADTcondition}
\end{align}
These two conditions coincide when the model parameters are correctly learned (i.e., the Nishimori condition), i.e., $\Delta \hat{c}_{\alpha} = \Delta c_{\alpha}$ for all $\alpha$.

\subsection{Spectral band}
We first consider the boundary $\lambda_{\mathrm{b}}$ of the spectral band. 
By applying the result derived by the cavity method in Ref.~\cite{NeriMetzPRL2016}, we have 
\begin{align}
|\lambda_{\mathrm{b}}| &= \frac{1}{q\sqrt{c}} \sqrt{ \sum_{\alpha>0} \frac{|\Delta \hat{c}_{\alpha}|^{2}}{P_{\alpha}} }. 
\label{lambda-b}
\end{align}
In terms of $\hat{x}_{\alpha}$, 
\begin{align}
|\lambda_{\mathrm{b}}| 
= \frac{\sqrt{c}}{q \Omega(1 - \Omega)} \sqrt{ \sum_{\alpha>0} P_{\alpha} |\hat{x}_{\alpha} - \Omega|^{2} }. \label{lambda-b-xalpha}
\end{align}
As shown in Appendix~\ref{MethodOfTypes}, this can also be derived as an upper bound of the spectral band by using the method of types. 
In the case of two equally sized modules with a simple community structure under the Nishimori condition, the condition $|\lambda_{2}(W)| |\lambda_{\mathrm{b}}| = 1$ yields 
\begin{align}
\sqrt{\sum_{\alpha>0} \frac{|\Delta c_{\alpha}|^{2}}{P_{\alpha}}} = 2 \sqrt{c}. \label{HeimlicherThreshold}
\end{align}

\begin{figure}[t]
 \begin{center}
   \includegraphics[width=0.8 \columnwidth]{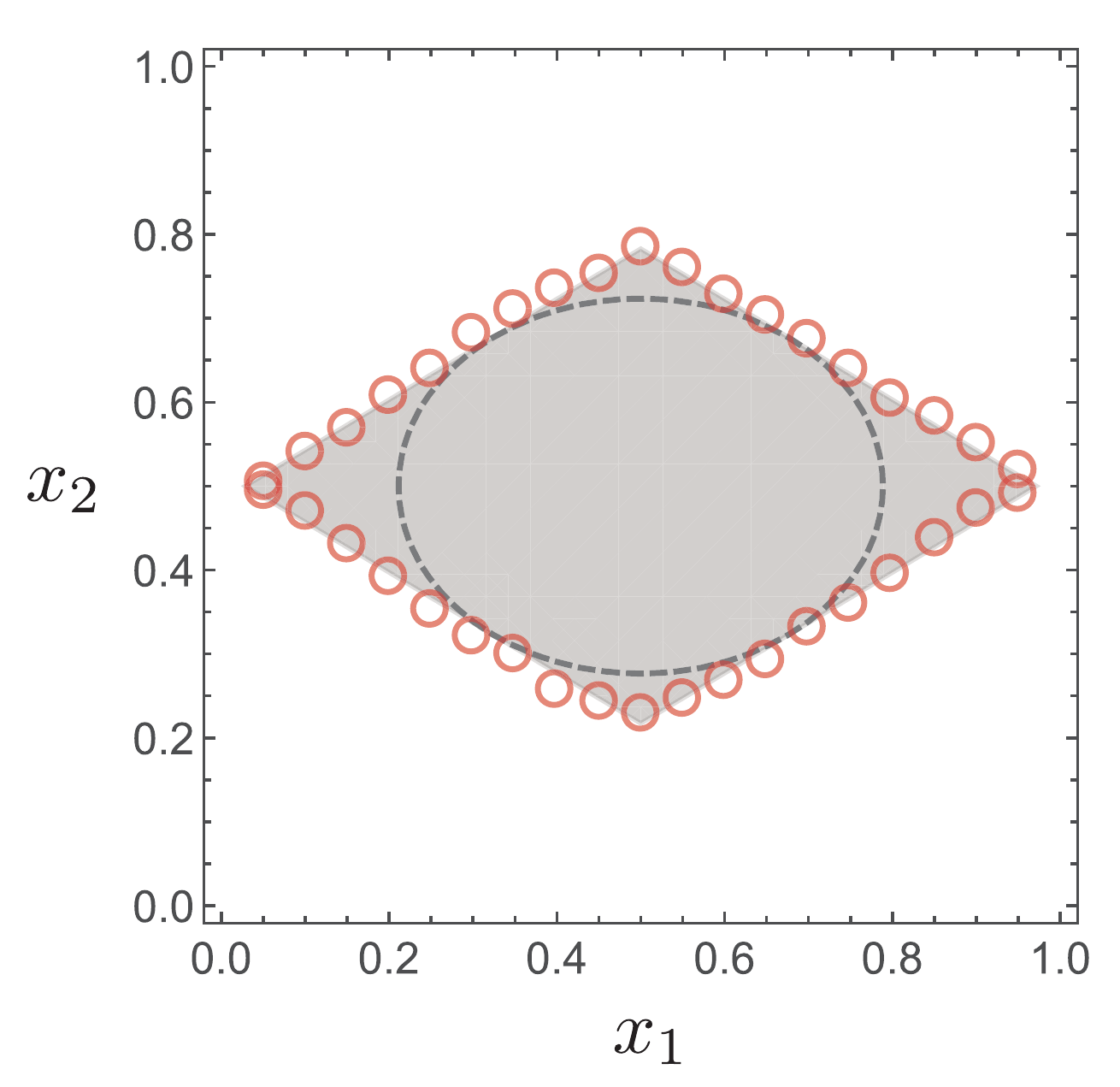}
 \end{center}
 \caption{
	(Color online) Detectability phase-diagram of the two equally sized ($q=2$) labeled SBM with two types of edges ($p=2$). 
	Each axis represents the normalized strength of the modular structure $x_{\alpha}$ ($\alpha \in \{1,2\}$). 
	The center of the diagram represents the uniform random graph, while the edges of the diagram represent the strongly modular graphs. 
	The size of the graph is $N=10,000$. 
	The average degree of each edge type is $c_{1} = 3$ and $c_{2} = 5$. 
	The shaded region represents the undetectable region, i.e., the region where the inferred module assignments by the EM algorithm are uncorrelated to the planted assignments. 
	The dashed ellipse represents the detectability threshold under the Nishimori condition. 
	(The ellipse becomes a circle when the average degrees of both edge types are equal.) 
	On the other hand, the shaded region represents the undetectable region of the EM algorithm, and its boundary is the algorithmic detectability threshold that we derived. 
	The circles represent the phase boundary obtained by the numerical experiment; each point represents the average over five samples. 
	}
 \label{EMPhaseDiagram-q2}
\end{figure}

This threshold is equal to the one derived in Ref.~\cite{Heimlicher2012}. 
As presented in Fig.~\ref{EMPhaseDiagram-q2}, however, the numerical experiment shows that the actual boundary where the EM algorithm fails (open circles) does not coincide with Eq.~(\ref{HeimlicherThreshold}) (dashed ellipse). 
Instead, the undetectable region can be well characterized by the shaded region; its boundary is the algorithmic detectability threshold that we will derive in Sec.~\ref{TwoCommunityPhaseDiagram}. 
In addition, we can confirm that the condition $|\lambda_{2}(W)| |\lambda_{\mathrm{b}}| = 1$ coincides with the threshold obtained in Ref.~\cite{KawamotoGMGM2017} under the Nishimori condition for the standard SBM with general modular structures.

\subsection{Isolated eigenvalues}
Next, we solve for the isolated eigenvalue $\lambda_{\mathrm{iso}}$. 
Given a graph, the eigenvalue equation of $B^{\prime}$ with respect to an eigenvector $v_{i \to j}$ with an eigenvalue $\lambda$ is 
\begin{align}
\lambda v_{i \to j} &= \sum_{k \in \partial i \backslash j} \left.\frac{\Delta c_{\alpha}}{q c_{\alpha}}\right|_{\alpha=A_{ik}} v_{k \to i}. \label{NBT-EV-1}
\end{align}
As the edge label $A_{ik}$ is stochastically determined by the planted module assignments, we denote the eigenvector element together with the planted module assignments $\sigma_{i}$ and $\sigma_{k}$ as $v^{(\sigma_{k}, \sigma_{i})}_{k \to i}$. 

Note that the nonbacktracking matrix is an oriented matrix, i.e., when $B_{i \to j, j \to k} = 1$, then $B_{j \to k, i \to j} = 0$.
According to Ref.~\cite{NeriMetzPRL2016}, the isolated eigenvalue of an oriented matrix can be obtained by solving the eigenvalue equation of the averaged quantities (Eq.~(S59) in Ref.~\cite{NeriMetzPRL2016}). 
Because the eigenvector statistics have dependency on the module assignment in the present case, we let $u^{\sigma} = \bracket{v^{\sigma}_{i \to j}}$ be the ensemble average of the eigenvector element with module assignment $\sigma$ with respect to vertex $i$. 
The isolated eigenvalue can be obtained by solving the equation for $u^{\sigma}$. 

Taking the ensemble average of the eigenvector elements and the configuration average, we have 
\begin{align}
\lambda u^{\sigma} &= 
\sum_{k=1}^{N-2} \sum_{\sigma^{\prime}} p(\sigma_{k} = \sigma^{\prime}) \mathbb{E}_{A,\sigma^{\prime}}\left[ \frac{\Delta c_{\alpha}}{q c_{\alpha}} \right] u^{\sigma^{\prime}}. 
\label{NBT-EV-2}
\end{align}
The probability $p(\sigma_{k} = \sigma^{\prime})$ that the neighboring vertex $k$ belongs to module $\sigma^{\prime}$ is given as $\gamma_{\sigma^{\prime}}$, and 
\begin{align}
\mathbb{E}_{A,\sigma^{\prime}}\left[ \frac{\Delta c_{\alpha}}{q c_{\alpha}} \right] 
&= \sum_{\alpha > 0} \frac{\Delta c_{\alpha}}{q c_{\alpha}} p(\alpha | \sigma, \sigma^{\prime}) \notag\\
&= \sum_{\alpha > 0} \frac{\Delta c_{\alpha}}{q c_{\alpha}} \frac{c^{\alpha}_{\sigma \sigma^{\prime}}}{N}
\end{align}
is the configuration average for given $\sigma^{\prime}$, and $p(\alpha | \sigma, \sigma^{\prime})$ is the probability that a neighboring vertex in module $\sigma^{\prime}$ is connected to a vertex in module $\sigma$ via an $\alpha$-edge. 

Making use of the fact that the module sizes are equal, we have 
\begin{align}
\lambda u^{\sigma} = \sum_{\sigma^{\prime}} J_{\sigma^{\prime} \sigma} u^{\sigma^{\prime}}, 
\hspace{10pt} J_{\sigma \sigma^{\prime}} &\equiv \frac{1}{q} \sum_{\alpha>0} c^{\alpha}_{\sigma,\sigma^{\prime}} \frac{\Delta \hat{c}_{\alpha}}{q c_{\alpha}}.  \label{NBT-EV-3}
\end{align}
The matrix $J$ can be expressed using $W$, and the eigenvalues of Eq.~(\ref{NBT-EV-3}) are written as follows.
\begin{align}
&\begin{cases}
\lambda_{+} = a \Delta J + q J_{\mathrm{out}}\\
\lambda_{-} = \lambda_{2}(W) \Delta J, 
\end{cases}
&\begin{cases}
\Delta J = \sum_{\alpha>0} \frac{\Delta c_{\alpha}}{q \sqrt{c_{\alpha}}} \frac{\Delta \hat{c}_{\alpha}}{q \sqrt{c_{\alpha}}} \\
J_{\mathrm{out}} = \frac{1}{q}\sum_{\alpha>0} c^{\alpha}_{\mathrm{out}} \frac{\Delta \hat{c}_{\alpha}}{q c_{\alpha}}.
\end{cases}\label{q2EigenvalueEq}
\end{align}
The eigenvector that corresponds to $\lambda_{+}$ in Eq.~(\ref{q2EigenvalueEq}) is proportional to a unit vector. 
Thus, $\lambda_{-}$ is the eigenvalue that we referred to $\lambda_{\mathrm{iso}}$, i.e., 
\begin{align}
\lambda_{\mathrm{iso}} &= \lambda_{2}(W) \sum_{\alpha>0} \frac{\Delta c_{\alpha}}{q \sqrt{c_{\alpha}}} \frac{\Delta \hat{c}_{\alpha}}{q \sqrt{c_{\alpha}}}. 
\label{lambda-iso}
\end{align}
In terms of $x_{\alpha}$ and $\hat{x}_{\alpha}$, 
\begin{align}
\lambda_{\mathrm{iso}} 
&= \frac{\lambda_{2}(W)}{\left[ q \Omega (1-\Omega)\right]^{2}} \sum_{\alpha>0} c_{\alpha} (x_{\alpha} - \Omega) (\hat{x}_{\alpha} - \Omega). 
\label{lambda-iso-x}
\end{align}

In summary, the algorithmic detectability threshold is determined by Eq.~(\ref{ADTcondition}) in which, $\lambda_{\mathrm{b}}$ and $\lambda_{\mathrm{iso}}$ are given by Eqs.~(\ref{lambda-b}) and (\ref{lambda-iso}), respectively.

\section{Detectability phase-diagrams}\label{Sec:DetectabilityPhaseDiagram}
In this section, we draw detectability phase-diagrams for some specific cases. 
Note that the algorithmic detectability threshold of the EM algorithm depends on the initial condition, i.e., we need to specify the initial estimates of the model parameters. 
As we will see for each example, the trajectory of the set of estimated model parameters is crucial to the geometry of the undetectable phase. 
In the following examples, we always set the number of edge types $p=2$.

\subsection{Community structure with $q=2$}\label{TwoCommunityPhaseDiagram}

\begin{figure*}[t!]
 \begin{center}
   \includegraphics[width=1.8\columnwidth]{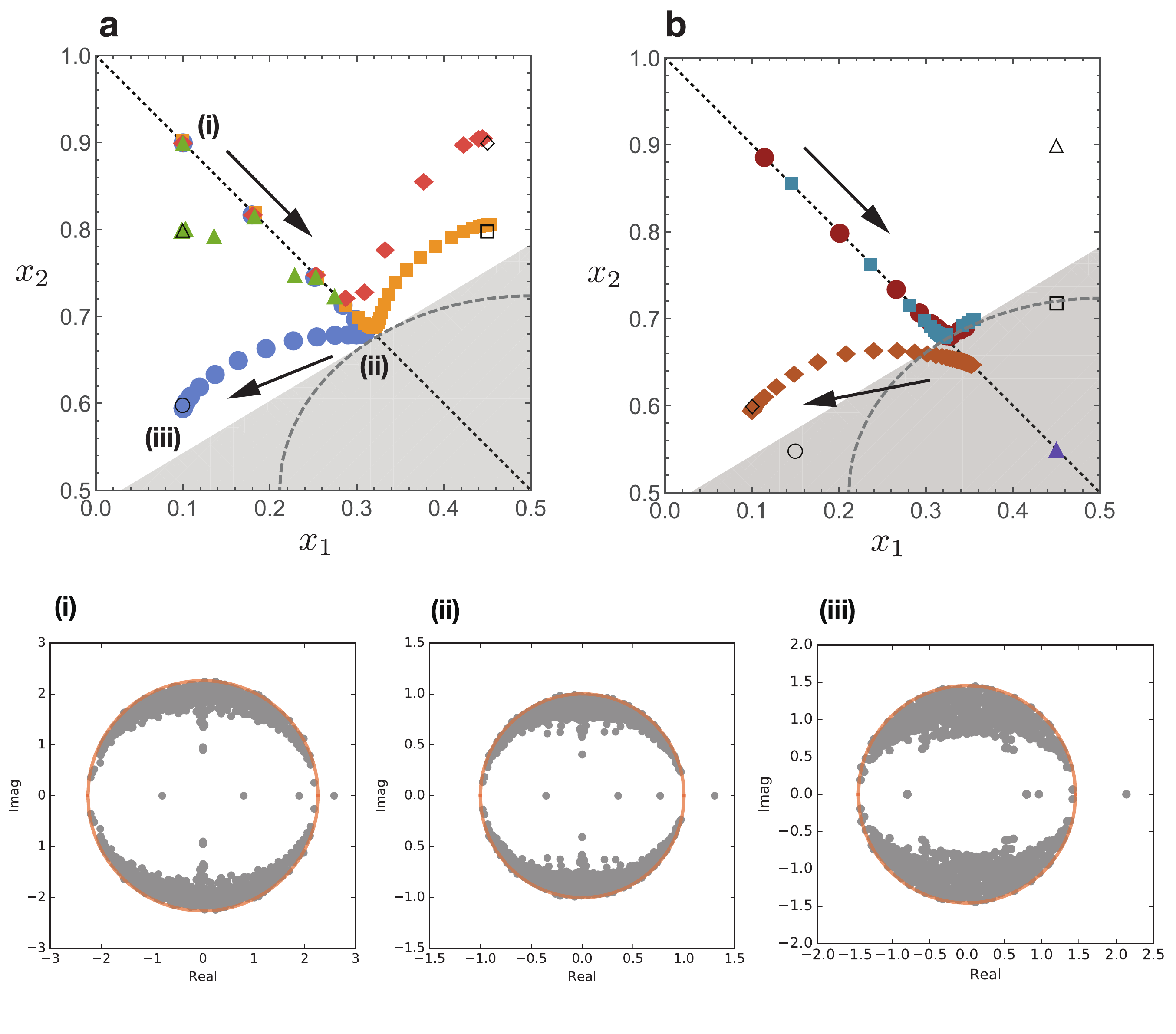} 
 \end{center}
 \caption{
	(Color online) 
	(Top) Trajectories of parameter learning based on the M-step of the EM algorithm for various planted values of $(x_{1}, x_{2})$. 
	This plot shows the upper-left region of the phase diagram shown in Fig.~\ref{EMPhaseDiagram-q2}, and we consider the same labeled SBM as in Fig.~\ref{EMPhaseDiagram-q2}. 
	The arrows show the directions in which the estimated parameters move. 
	({\bf a}) The trajectories of the estimates $(\hat{x}_{1}, \hat{x}_{2})$ for the case that the planted values $(x_{1}, x_{2})$ (shown in open symbols) are in the detectable region. 
	The trajectory for the graph with the planted value $(x_{1}, x_{2}) = (0.1, 0.6)$ is represented by blue circles; 
	$(x_{1}, x_{2}) = (0.45, 0.8)$ is represented by yellow squares; 
	$(x_{1}, x_{2}) = (0.45, 0.9)$ is represented by red diamonds; and 
	$(x_{1}, x_{2}) = (0.1, 0.8)$ is represented by green triangles, respectively. 
	In all cases, the initial estimate is set as $(\hat{x}_{1}, \hat{x}_{2}) = (0.1, 0.9)$. 
	The dotted line represents the line with slope $-1$. 
	({\bf b}) The trajectories of the estimates $(\hat{x}_{1}, \hat{x}_{2})$ in other cases. 
	The trajectories with the planted values $(x_{1}, x_{2}) = (0.15, 0.55)$ (red circles) and $(x_{1}, x_{2}) = (0.45, 0.72)$ (cyan squares) are the cases in which the planted values (shown in open symbols) are in the undetectable region though the initial estimates are set as $(\hat{x}_{1}, \hat{x}_{2}) = (0.1, 0.9)$. 
	The trajectories with the planted values $(x_{1}, x_{2}) = (0.1, 0.6)$ (orange diamonds) and $(x_{1}, x_{2}) = (0.45, 0.9)$ (purple triangles) are the cases in which $(\hat{x}_{1}, \hat{x}_{2})$ is initially located in the undetectable region ($(\hat{x}_{1}, \hat{x}_{2}) = (0.45, 0.55)$) though the planted values are in the detectable region. 
	(Bottom) Spectra of the weighted nonbacktracking matrix $B^{\prime}$ in the complex plane with $N=500$ corresponding to ({\bf i}) $(\hat{x}_{1}, \hat{x}_{2}) = (0.1, 0.9)$, ({\bf ii}) $(\hat{x}_{1}, \hat{x}_{2}) = (0.323, 0.677)$, and ({\bf iii}) $(\hat{x}_{1}, \hat{x}_{2}) = (0.1, 0.6)$. 
	The solid line (red) represents the circle with radius $|\lambda_{\mathrm{b}}|$. 
	}
 \label{EMtrajectories-q2}
\end{figure*}

We first derive the phase boundary of the shaded region in Fig.~\ref{EMPhaseDiagram-q2}. 
This is a case of two equally sized modules ($q=2$), and $W$ is equal to the identity matrix (i.e., $|\lambda_{2}(W)| = 1$); this is often referred to as the symmetric SBM. 
Figure \ref{EMtrajectories-q2}a shows the trajectories of the estimate $(\hat{x}_{1},\hat{x}_{2})$ for various instances of the labeled SBMs. 
All the planted model parameters are in the detectable region. 

We set the initial estimate $(\hat{x}_{1},\hat{x}_{2})$ nearly at the corner of the $(x_{1}, x_{2})$-plane, $(0.1, 0.9)$. 
An example of the corresponding spectrum of the weighted nonbacktracking matrix $B^{\prime}$ is shown in Fig.~\ref{EMtrajectories-q2}(i); upon setting the initial condition as shown, the boundary of the spectral band exceeds 1.  
As described in Secs.~\ref{Sec:MstepTransientDynamics} and \ref{InstabilityOfFactorizedState}, $\hat{x}_{\alpha}$ of each $\alpha$ is attracted toward the point of the uniform graph at equal rates until it satisfies the condition $|\lambda_{\mathrm{b}}| = 1$ [Fig.~\ref{EMtrajectories-q2}(ii)], or equivalently,  $|\hat{x}_{\alpha} - 1/2| = (2\sqrt{c})^{-1}$ for both $\alpha$. 
Given these estimates, the condition $|\lambda_{\mathrm{iso}}|=1$ yields 
\begin{align}
\sum^{p}_{\alpha=1} P_{\alpha} \left| x_{\alpha} - \frac{1}{2} \right| = \frac{1}{2 \sqrt{c}}. \label{AlgoDetectabilitySymInit}
\end{align} 
This is the boundary of the shaded region in Fig.~\ref{EMPhaseDiagram-q2}. 
In terms of $\Delta c_{\alpha}$, Eq.~(\ref{AlgoDetectabilitySymInit}) is $\sum_{\alpha>0} |\Delta c_{\alpha}| = 2\sqrt{c}$. 

Thereafter, when the graph is in the detectable region, $(\hat{x}_{1},\hat{x}_{2})$ moves to the planted value. 
The spectrum of $B^{\prime}$ is shown in Fig.~\ref{EMtrajectories-q2}(iii). 
On the other hand, as shown in Fig.~\ref{EMtrajectories-q2}b (circles and squares), $(\hat{x}_{1},\hat{x}_{2})$ does not reach the planted value in the undetectable region; provided that $(\hat{x}_{1},\hat{x}_{2})$ is initially located in the detectable region, it gets stuck when $|\lambda_{\mathrm{b}}| = 1$ is satisfied. 

A value of model parameter that exhibits a strong modular structure is empirically known as a better choice for the initial model parameter. 
Indeed, when $(\hat{x}_{1},\hat{x}_{2})$ is initially located deep in the undetectable region, the estimate does not move at all [triangles in Fig.~\ref{EMtrajectories-q2}b]. 
This can be understood as the algorithmic detectability threshold; the condition $|\lambda_{2}(W)| |\lambda_{\mathrm{b}}| \le 1$ is already satisfied at the beginning of the algorithm, and there is no isolated eigenvalue of $B^{\prime}$ that satisfies $|\lambda_{2}(W)| |\lambda_{\mathrm{iso}}| > 1$. 
Similarly, although $(\hat{x}_{1},\hat{x}_{2})$ is initially located in the undetectable region, when $|\lambda_{2}(W)| |\lambda_{\mathrm{iso}}| > 1$ is satisfied, then the estimate moves successfully to the planted value [diamonds in Fig.~\ref{EMtrajectories-q2}b].

In the standard SBM, as long as the initial estimate of the model parameters is in the detectable region, we can confirm that there is no distinction between the algorithmic threshold and the threshold under the Nishimori condition; the undetectable region is given by $|\Delta c_{1}| < q \sqrt{c_{1}}$ in the case of a binary label $\alpha \in \{0,1\}$. 
This is consistent with Theorem 3 of Ref.~\cite{Mossel2015} that the model parameters are asymptotically learnable for $q=2$. 

The dependence of the initial estimate of the model parameters was also examined numerically in Ref.~\cite{Decelle2011a} for the standard SBM. 
If we substitute the value corresponding to the critical point $\epsilon_{\ell}$ in Fig.~5 (left) in Ref.~\cite{Decelle2011a} into Eq.~(\ref{lambda-iso}), we obtain $|\lambda_{\mathrm{iso}}| = 1$. In addition, according to Eq.~(\ref{lambda-b}), this result corresponds to the case where we start the algorithm with $|\lambda_{\mathrm{b}}|<1$; thus, the critical point $\epsilon_{\ell}$ was actually the algorithmic detectability threshold.

\subsection{Community structure with $q=3$}\label{ThreeCommunityPhaseDiagram}
\begin{figure*}[ht!]
 \begin{center}
   \includegraphics[width=2\columnwidth]{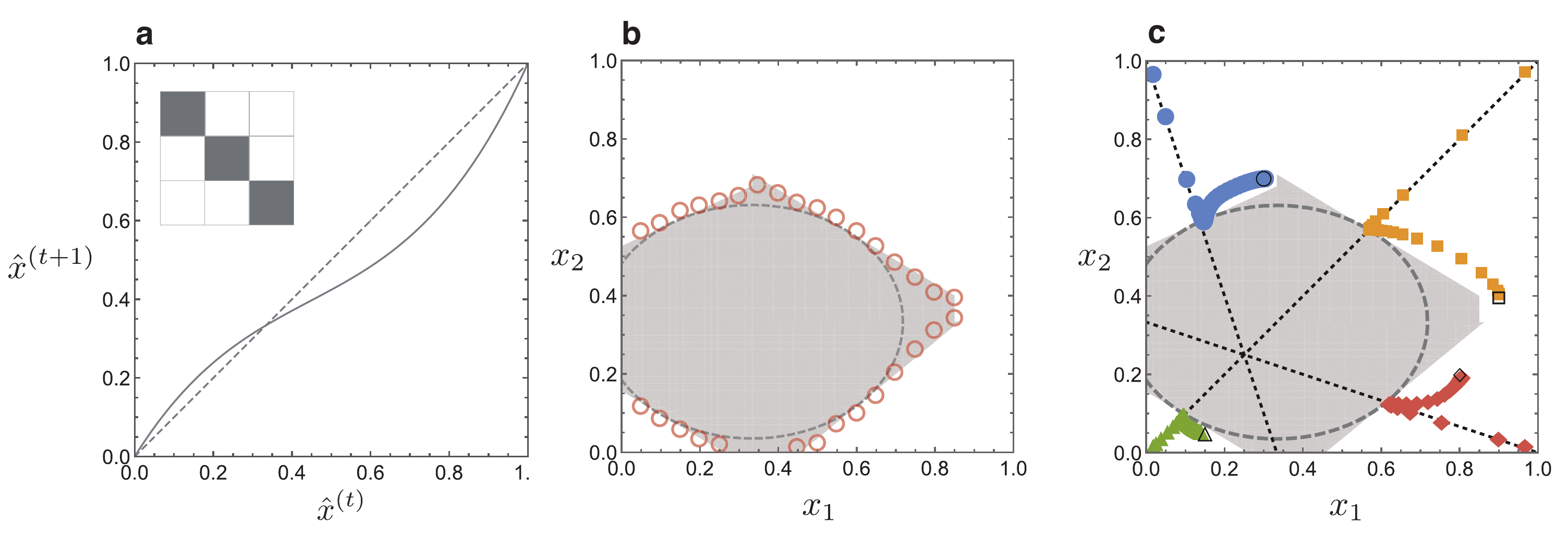}
 \end{center}
 \caption{
	(Color online) Behavior of the EM algorithm in the simple community structure with three modules. 
	The average degree of each edge-type is $c_{1} = 3$ and $c_{2} = 5$.
	We consider two types of edges ($p=2$). 
	({\bf a}) Second-order expansion of the right-hand side of Eq.~(\ref{xUpdate2}) with $\bracket{\xi_{ij}} = 0$; the dashed line represents $\hat{x}^{(t+1)} = \hat{x}^{(t)}$. 
	In the inset, $W$ matrix where $c_{\mathrm{in}}$ and $c_{\mathrm{out}}$ are represented by black and white, respectively. 
	({\bf b}) Detectability phase-diagram. The dashed ellipse, shaded region, and circles represent the detectability threshold under the Nishimori condition, undetectable region of the EM algorithm, and results of the numerical experiments (the average is taken over five samples), respectively as shown in Fig.~\ref{EMPhaseDiagram-q2}. 
	$(x_{1},x_{2}) = (1/3,1/3)$ represents the point of the uniform random graph. 
	({\bf c}) Trajectories of $(\hat{x}_{1}, \hat{x}_{2})$ with various initial values in the phase space. 
	The trajectory for the graph with the planted value $(x_{1}, x_{2}) = (0.3, 0.7)$ corresponds to the blue circles; 
	$(x_{1}, x_{2}) = (0.9, 0.4)$ corresponds to the yellow squares; 
	$(x_{1}, x_{2}) = (0.8, 0.2)$ corresponds to the red diamonds; and 
	$(x_{1}, x_{2}) = (0.15, 0.05)$ corresponds to the green triangles. 
	As in Fig.~\ref{EMtrajectories-q2}, the planted values are shown by open symbols. 
	The dashed lines represent the lines with slopes $-1/3$, $1$, and $-3$. 
	For the numerical experiments, we use the labeled SBMs with $N=15,000$. 
	}
 \label{EMPhaseDiagram-q3}
%
 \begin{center}
   \includegraphics[width=1.8\columnwidth]{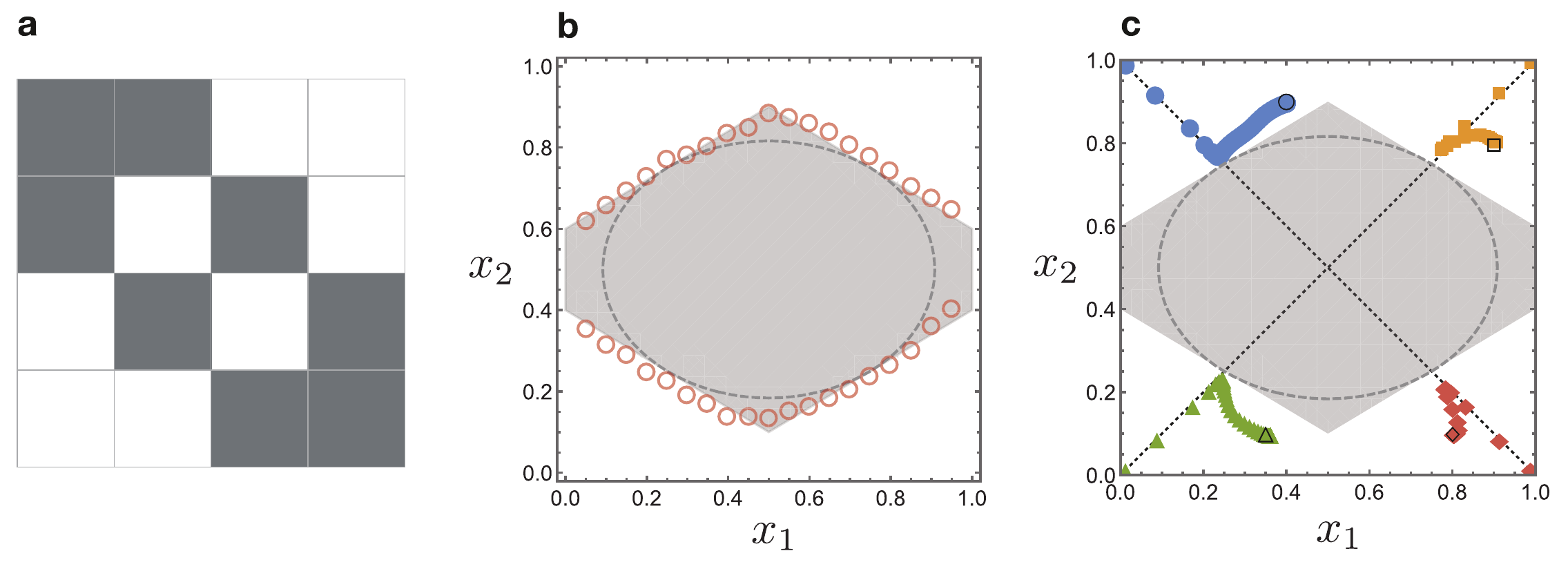}
 \end{center}
 \caption{
	(Color online) 
	Behavior of the EM algorithm in a noncommunity structure with two modules.
	The average degree of each edge-type is $c_{1} = 3$ and $c_{2} = 5$.
	({\bf a}) $W$ matrix where $c_{\mathrm{in}}$ and $c_{\mathrm{out}}$ are represented by black and white, respectively. We consider two types of edges ($p=2$). 
	({\bf b}) Detectability phase-diagram. The dashed ellipse, shaded region, and circles represent the detectability threshold under the Nishimori condition, undetectable region of the EM algorithm, and results of the numerical experiments (the average is taken over five samples), respectively, as in Fig.~\ref{EMPhaseDiagram-q2}. 
	$(x_{1},x_{2}) = (1/2,1/2)$ represents the point of the uniform random graph. 
	({\bf c}) Trajectories of $\hat{x}_{\alpha}$ with various initial values in the phase space. 
	The trajectory for the graph with the planted value $(x_{1}, x_{2}) = (0.4, 0.9)$ corresponds to the blue circles; 
	$(x_{1}, x_{2}) = (0.9, 0.8)$ corresponds to the yellow squares; 
	$(x_{1}, x_{2}) = (0.8, 0.1)$ corresponds to the red diamonds; and 
	$(x_{1}, x_{2}) = (0.35, 0.1)$ corresponds to the green triangles.
	As in Fig.~\ref{EMtrajectories-q2}, the planted values are shown by open symbols. 
	The dashed lines represent lines with slopes $-1$ and $1$. 
	For the numerical experiments, we use the labeled SBMs with $N=4,000$. 
	}
 \label{EMPhaseDiagram-GMGM}
\end{figure*}

When $\Omega = 1/2$ as in Sec.~\ref{TwoCommunityPhaseDiagram}, we observed that the rate of attraction of $\hat{x}_{\alpha}$ toward the point of the uniform random graph is equal (or almost equal) for all $\alpha$. 
As an example showing that this is not the case, let us consider the simple community structure with three equally sized modules. 
The results are shown in Fig.~\ref{EMPhaseDiagram-q3}. 
The value of $\Omega$ is $1/3$, and the update equation of $(\hat{x}_{1}, \hat{x}_{2})$ is no longer symmetric with respect to $\alpha$. 
As a consequence, the rate of attraction in the transient dynamics of $\hat{x}_{\alpha}$ differs depending on whether we set $\hat{x}_{\alpha} < \Omega$ or $\hat{x}_{\alpha} > \Omega$ at the beginning of the algorithm. 

Once we determine the point where the estimate $\hat{\ket{x}}$ hits the boundary of the spectral band, we can derive the detectability threshold by $ |\lambda_{\mathrm{iso}}| = 1$. 
Note that the resulting detectability phase-diagram has more detectable region in the upper-right side than the lower-left side; this reflects the fact that the edges within a module are more informative than the edges between modules when $q>2$.

\subsection{Non-community structure}\label{FourGMGMPhaseDiagram}
In the previous examples, we focused on the cases with $|\lambda_{2}(W)| = 1$. 
In this section, we describe the case of a noncommunity structure such that the $|\lambda_{2}(W)|$-dependence of the detectability threshold is actually observed. 
We consider the graph with the matrix $W$ shown in Fig.~\ref{EMPhaseDiagram-GMGM}a, i.e., $q=4$, $\Omega = 1/2$, $\lambda_{2}(W) = \sqrt{2}$. 
As in Sec.~\ref{TwoCommunityPhaseDiagram}, we set the initial condition that $|\hat{x}_{\alpha} - 1/2| = \text{const.}$ for any $\alpha$. 
All estimates $\hat{x}_{\alpha}$ are attracted toward the center of the phase space at equal rates until they satisfy the condition $|\lambda_{2}(W)| |\lambda_{\mathrm{b}}| = 1$, which yields $|\hat{x}_{\alpha} - 1/2| = 1/\sqrt{2c}$. 
Given these estimates, the condition $|\lambda_{2}(W)| |\lambda_{\mathrm{iso}}| = 1$ yields 
\begin{align}
\sum^{p}_{\alpha=1} P_{\alpha} \left| x_{\alpha} - \frac{1}{2} \right| = \frac{1}{\sqrt{2c}}. 
\end{align}
The detectability phase-diagram and the trajectories of the estimate $(\hat{x}_{1}, \hat{x}_{2})$ are shown in Figs.~\ref{EMPhaseDiagram-GMGM}b and \ref{EMPhaseDiagram-GMGM}c, respectively. 

To obtain the results shown in Fig.~\ref{EMPhaseDiagram-GMGM}, although we could have used the EM algorithm with the restricted affinity matrix in Eq.~(\ref{affinityGMGM}), we used the affinity matrix of full degrees of freedom instead. 
We can confirm that the boundary of the detectability phase-diagram is still very accurate, and thereby, the restriction of the affinity matrix that we imposed for analytical tractability does not have a crucial effect after all.

\section{Algorithmic infeasibility}\label{Sec:AlgoInfeasibility}
In this section, we focus on the case described in Sec.~\ref{TwoCommunityPhaseDiagram}; for this case, we discuss the physical consequence of the distinction between the algorithmic detectability threshold, Eq.~(\ref{AlgoDetectabilitySymInit}), and the detectability threshold under the Nishimori condition, Eq.~(\ref{HeimlicherThreshold}). 

Suppose that we have an instance of the standard SBM, whose edges indicate the assortative structure. 
However, the planted structure is undetectable by the EM algorithm because the graph is too sparse. 
To improve this situation, we can add some edges of a new type to the existing graph. 

Because we obtain more information about the planted structure by adding these edges of a new type, in principle, the structure is more likely to be detectable. 
Then, the question arises whether such a prescription always strengthens the detectability and whether is it even better to introduce yet another type of edges. 
In practice, the above statements are not true. 
An algorithm does not always perform better because it becomes more difficult to learn higher dimensional model parameters. 
In other words, excessively higher-order information will be, at some point, algorithmically infeasible to extract. 

\begin{figure}[t]
 \begin{center}
   \includegraphics[width=0.85\columnwidth]{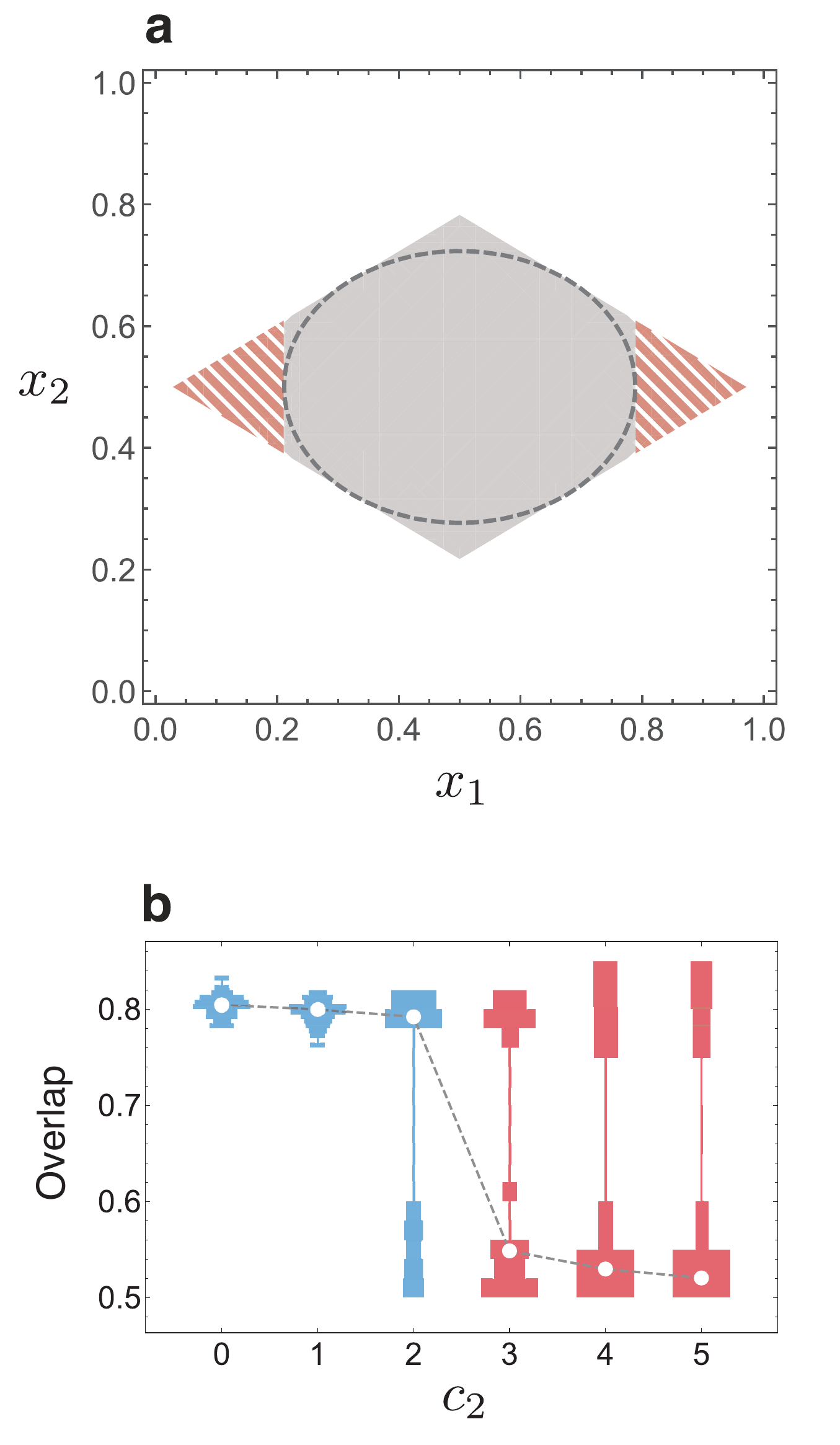}
 \end{center}
 \caption{
	(Color online) ({\bf a}) Same detectability phase-diagram as in Fig.~\ref{EMPhaseDiagram-q2}. 
	The striped region represents the undetectable region that becomes detectable when the edges of $\alpha=2$ are discarded. 
	({\bf b}) Vertical histograms of the overlap distribution. 
	The same labeled SBM as that in Fig.~\ref{InfeasibilityPhaseDiagram-q2}a is considered, and the instances with $N=10,000$, $c_{1}=3$, and $(x_{1}, x_{2}) = (0.85,0.45)$ are generated. 
	The histograms of various values of $c_{2}$ are horizontally aligned (30 samples for each histogram). 
	The ones in the detectable phase, i.e., $c_{2} \le 2$, are indicated in blue, while the ones in the undetectable phase, i.e., $c_{2} \ge 3$, are in red. 
	The white points (connected via dashed lines) indicate the medians of overlaps. 
	The population of the success and failure changes at the critical value that we estimated. 
	}
 \label{InfeasibilityPhaseDiagram-q2}
\end{figure}

This infeasibility can also be explained in the reverse way: There are cases where discarding edges of one type improves the detectability. 
To observe this behavior, we identify a region of the undetectable phase in the phase diagram of $p=2$ where the corresponding graph becomes detectable upon discarding the edges of $\alpha = 2$. 
Because the undetectable region of $\alpha = 1$ is given by 
\begin{align}
\left|x_{1} - \frac{1}{2}\right| < \frac{1}{2\sqrt{3}}, \label{alpha1Threshold}
\end{align}
we can readily see that the striped region in Fig.~\ref{InfeasibilityPhaseDiagram-q2}a is the phase where algorithmic infeasibility can be observed. 

The improvement of the performance is confirmed in Fig.~\ref{InfeasibilityPhaseDiagram-q2}b. 
This is a set of vertical histograms with respect to the \textit{overlap}, the fraction of correctly classified vertices. 
Each histogram shows the distribution of overlaps for the given average degree $c_{2}$ of $\alpha=2$. 
As we increase $c_{2}$, at some point, the graph will enter the undetectable region; according to Eq.~(\ref{AlgoDetectabilitySymInit}), the critical value is $c_{2} \approx 2.54$. 
Indeed, the numerical experiment shows that the overlaps tend to be high for $c_{2} \le 2$ (blue) and tend to be close to $0.5$, i.e., not better than chance, for $c_{2} \ge 3$ (red). 
Note that our analytical results are of $N \to \infty$. 
Thus, there is a chance to retrieve the information of the planted modules even for $c_{2} \ge 3$ because of the finite-size effect and vice versa. 
Note also that if we make the average degree $c_{2}$ even larger ($c_{2} \gtrsim 55.46$ in the current case), the planted modules eventually become detectable again. 

Importantly, the detectability threshold of $\alpha = 1$, Eq.~(\ref{alpha1Threshold}), is tangent to the detectability threshold under the Nishimori condition (dashed ellipse). 
Therefore, the emergence of the phase that exhibits the algorithmic infeasibility is a consequence of the algorithmic detectability threshold. 

\section{Summary and Discussion}\label{Sec:SummaryDiscussion}
In this study, we derived the algorithmic detectability threshold of the labeled SBM by using the EM algorithm. 
Although we restricted the parameters to enable analytical calculation, our result is applicable to more than two modules, arbitrary number of edge types $p$, and general modular structures. 
Our result offers another aspect to the detectability threshold in statistical inference. 
Although BP is known to achieve the theoretical limit of detectability in some situations, the EM algorithm that uses BP in its E-step cannot achieve that limit unless a special initial condition is chosen. 
This difference stems from the learnability of the model parameters. 
This is also a crucial difference between graph clustering and the tree reconstruction problem. 
Although they are closely related with regard to detectability, in the tree reconstruction problem, the Markov transition matrix is given as an input, i.e., the problem of learnability is absent. 

Note that the result obtained considering the instability of the factorized state is not a proof of the detectability threshold. (See e.g., Refs.~\cite{Mossel2015,Massoulie2014,Lelarge2015} for a detailed discussion.)
Nevertheless, as we observed in Sec.~\ref{Sec:DetectabilityPhaseDiagram}, our analysis predicts the behavior of numerical experiments very accurately. 
We also did not analyze the emergence of the so-called hard phase that typically appears when the number of modules $q$ is large. 
The analysis in this study deals with the dynamics of the model parameter for a given graph. On the other hand, reports on sample-averaged dynamics of the EM algorithm are available in literature \cite{Inoue2001,InoueJphysA03,TanakaJphysA07,Kataoka2012}, mainly in the context of image restoration. 

In general, the phase boundary that we derived (e.g., the boundary of the shaded region in Fig.~\ref{EMPhaseDiagram-q2}) is a simplex that is tangent to the detectability threshold under the Nishimori condition (e.g., the dashed ellipse in Fig.~\ref{EMPhaseDiagram-q2}), and the tangent point depends on the initial estimate of $\hat{x}_{\alpha}$. 
Note that the EM algorithm encounters this limitation when the critical conditions that we derived, as shown in the Appendix, are met in the transient regime. 
In other words, for example, when the planted sizes of modules are very different, the transient regime will be too short, and the M-step trajectory becomes very complicated. 
This implies that it is usually impossible to track the behavior of the EM algorithm, and it is rather surprising that there is a class of the SBM (and it is not too restricted) for which we can derive the algorithmic detectability threshold analytically. 

Although our analysis here deals with graph clustering using the SBM, we expect that the strategy here can be applied to other models, particularly in machine learning. 
We hope that the analysis of algorithms utilizing the geometry of the phase diagram offers deeper insights to various kinds of problems.

\section*{acknowledgments}
\textit{Acknowledgments}--- The author thanks Yoshiyuki Kabashima, Tomoyuki Obuchi, and Jean-Gabriel Young for fruitful discussions and valuable comments. This work was supported by the New Energy and Industrial Technology Development Organization (NEDO).

\appendix 

\section{Rate of attraction in the M-step transient dynamics}
The example given in the main text shows the equal rate of attraction for each $\hat{x}_{\alpha}$ toward the point of the uniform graph. 
This implies that higher order moments $\bracket{\xi_{ij}^{k}}$ ($k>2$) are absent or negligible. 
This can be explained as follows. 
Suppose we set the initial values of the model parameters $\hat{x}_{1} = \varepsilon$ and $\hat{x}_{2} = 1- \varepsilon$. In general, their rates of attraction toward the point of the uniform random graph are not equal. However, when $\Omega = 1/2$ and the higher-order moments of $\xi_{ij}$ are zero, the update equation (\ref{xUpdate2}) is invariant under the transform $\hat{x}^{(t+1)}\to 1-\hat{x}^{(t+1)}$ and $\hat{x}^{(t)}\to 1-\hat{x}^{(t)}$, indicating that they are attracted at equal rates.


\section{Possibility to improve the learnability}
One might wonder whether we can improve the learnability of the model parameters by tuning the EM algorithm. 
For example, the algorithm may achieve better learnability if we can control the speed of model-parameter learning. 

As a matter of fact, our implementation of the EM algorithm is not a precise implementation. 
In principle, the EM algorithm requires iterations until convergence for every E-step, and then, we stop the algorithm when the M-step converges. 
Here, we need to introduce a convergence criterion for each of iteration and a cutoff for the number of iterations. 
Note, however, that unless the convergence in the E-step is extremely quick, this legitimate implementation of the EM algorithm with the large iteration-cutoff will be very time consuming. 
Thus, it is reasonable to set a small value of the iteration-cutoff. 
Here we set this cutoff as 1. In other words, the model parameters are updated for every sweep of BP. 
Moreover, we do not set a convergence criterion for the M-step. Instead, we use the convergence of the E-step with respect to the previous update; thus, we only need one convergence criterion. 

Although we also tested the performance of the legitimate implementations that have the iteration-cutoffs of 10 and 100, we did not observe any improvement. 
The insensitivity to the implementation detail can be interpreted from the analysis in Appendix~\ref{PsiTransientDynamics}; even if we start from a nonuniform distribution of the module-assignment estimate that may be positively correlated to the planted assignment, the transient dynamics randomizes that distribution. 

It is also common to introduce a parameter called the learning rate $\eta$ ($<1$) to control the update rate of model parameters. 
The update equation of the estimate $\hat{x}_{\eta}$ with the learning rate $\eta$ is written as follows. 
\begin{align}
\hat{x}^{(t+1)}_{\eta} &= (1-\eta) \hat{x}^{(t+1)}_{0} + \eta \, \hat{x}^{(t+1)}_{1} \notag\\
&= \hat{x}^{(t)}_{\eta} \left[ 1 + \eta \left( \bracket{ \frac{1+\xi_{ij}}{1 + \frac{\hat{x}^{(t)}-\Omega}{1-\Omega}\xi_{ij} } } - 1 \right) \right].
\end{align}
While the learning rate slows down the update of $\hat{x}$ by definition, it does not alter the fixed points. 
As far as we tested, again, we did not observe any improvement in performance. 
We therefore conclude that the algorithmic detectability threshold we derived can hardly be improved by tuning the implementation.

\section{Derivation of the boundary of the spectral band}\label{MethodOfTypes}
Here, we derive an upper bound of the spectral band of the weighted nonbacktracking matrix $B^{\prime}$. 
This is an extension of the derivation in Ref.~\cite{Krzakala2013} for an unlabeled graph. 
The bound here coincides with the threshold obtained in Ref.~\cite{Heimlicher2012}, as derived using the large deviation technique; however, we use the method of types \cite{CoverThomas}. 

As discussed in Ref.~\cite{Krzakala2013}, we consider the following relation with respect to the eigenvalues $\{\lambda_{\ell}\}$ of the nonbacktracking matrix. 
\begin{align}
\sum_{\ell=1}^{2L} |\lambda_{\ell}|^{2d} 
&\le \mathrm{tr} B^{d} \left( B^{d} \right)^{\top} \notag\\
&= \sum_{v_{i \to j}} \sum_{v_{w \to x}} \left\lvert v_{i \to j} B^{d} v_{w \to x} \right\rvert^{2} \label{SpectralBand1}
\end{align}
In the case of the unweighted edges, the sum over $v_{w \to x}$ can be interpreted as the number of the nonbacktracking paths that reach edge $(i,j)$ by exactly $d$ steps. 
When the graph is treelike, it is approximately $c^{d}$. 

We generalize the above argument to the weighted nonbacktracking matrix $B^{\prime}$. 
First, given the population $P = \{P_{1}, \cdots, P_{p}\}$ of the number of labeled edges, we denote the fraction of paths that have the empirical distribution, i.e., the type, $\tilde{P} = \{\tilde{P}_{1}, \cdots, \tilde{P}_{p}\} \in \mathcal{P}$, along a path of distance $d$ as $\rho_{d}[\tilde{P}:P]$. 
When $d \gg 1$, we can approximately write it as 
\begin{align}
\rho_{d}[\tilde{P}:P] 
&= 
\begin{pmatrix}
& d \\
d\tilde{P_{1}}&\cdots&d\tilde{P_{p}} 
\end{pmatrix}
\prod_{\alpha>0} P_{\alpha}^{d\tilde{P}_{\alpha}} \notag\\
&\sim \mathrm{e}^{-dD(\tilde{P}||P)}, 
\end{align}
where $D(\tilde{P}||P)$ is the Kullback--Leibler divergence of distributions $\tilde{P}$ and $P$. 
Using this quantity, we can express a part of Eq.~(\ref{SpectralBand1}) as 
\begin{align}
&\sum_{v_{w \to x}} \left\lvert v_{i \to j} B^{\prime d} v_{w \to x} \right\rvert^{2} \notag\\
&= c^{d} \sum_{\tilde{P} \in \mathcal{P}} \rho[\tilde{P}:P] \prod_{\alpha>0} \left| \frac{\Delta \hat{c}_{\alpha}}{q c_{\alpha}} \right|^{2d \tilde{P}_{\alpha}} \notag\\ 
&= \left( \frac{\sum_{\alpha>0} |\Delta \hat{c}_{\alpha}|}{q\sqrt{c}} \right)^{2d} \sum_{\tilde{P} \in \mathcal{P}} \mathrm{e}^{d \left( D(\tilde{P}||P) - 2D(\tilde{P}||Q) \right)},  \label{SpectralBand2}
\end{align}
where we defined $Q_{\alpha} = |\Delta \hat{c}_{\alpha}|/\sum_{\alpha>0} |\Delta \hat{c}_{\alpha}|$. 
In the limit $d \to \infty$, the law of large numbers ensures 
\begin{align}
\sum_{\tilde{P} \in \mathcal{P}} \mathrm{e}^{d \left( D(\tilde{P}||P) - 2D(\tilde{P}||Q) \right)}
\sim \left(\sum_{\alpha>0} \frac{Q_{\alpha}^{2}}{P_{\alpha}}\right)^{d}.  \label{SpectralBand3}
\end{align}
Hence, from Eqs.~(\ref{SpectralBand1}), (\ref{SpectralBand2}), and (\ref{SpectralBand3}), we have 
\begin{align}
\frac{1}{2L} \sum_{\ell=1}^{2L} |\lambda_{\ell}|^{2d} 
&\lesssim \left[ \frac{1}{q^{2}c} \sum_{\alpha>0} \frac{|\Delta \hat{c}_{\alpha}|^{2}}{P_{\alpha}} \right]^{d}. 
\end{align}
Therefore, $\lambda^{2}_{\mathrm{b}}$ is estimated as 
\begin{align}
\lambda^{2}_{\mathrm{b}} &\lesssim \frac{1}{q^{2}c} \sum_{\alpha>0} \frac{|\Delta \hat{c}_{\alpha}|^{2}}{P_{\alpha}}, 
\end{align}
and this is equal to Eq.~(\ref{lambda-b}). 
Note that the fact that we consider the general modular structure does not comes into play here owing to the assumption that the expected degree of each vertex does not depend on the module to which it belongs.

\section{Randomization of the module assignment distributions}\label{PsiTransientDynamics}
Here, we show that the module assignment distributions are randomized during the transient dynamics. 
We present three cases of the labeled SBM as examples: the simple community structure with $q=2$ and $q=3$, and the general modular structure of Fig.~\ref{EMPhaseDiagram-GMGM}a. 
In all cases, we set the number of labels $p=2$ and choose a corner of the phase space for the initial estimate of each $\hat{x}_{\alpha}$. 

\begin{figure*}[t]
 \begin{center}
   \includegraphics[width=1.8 \columnwidth]{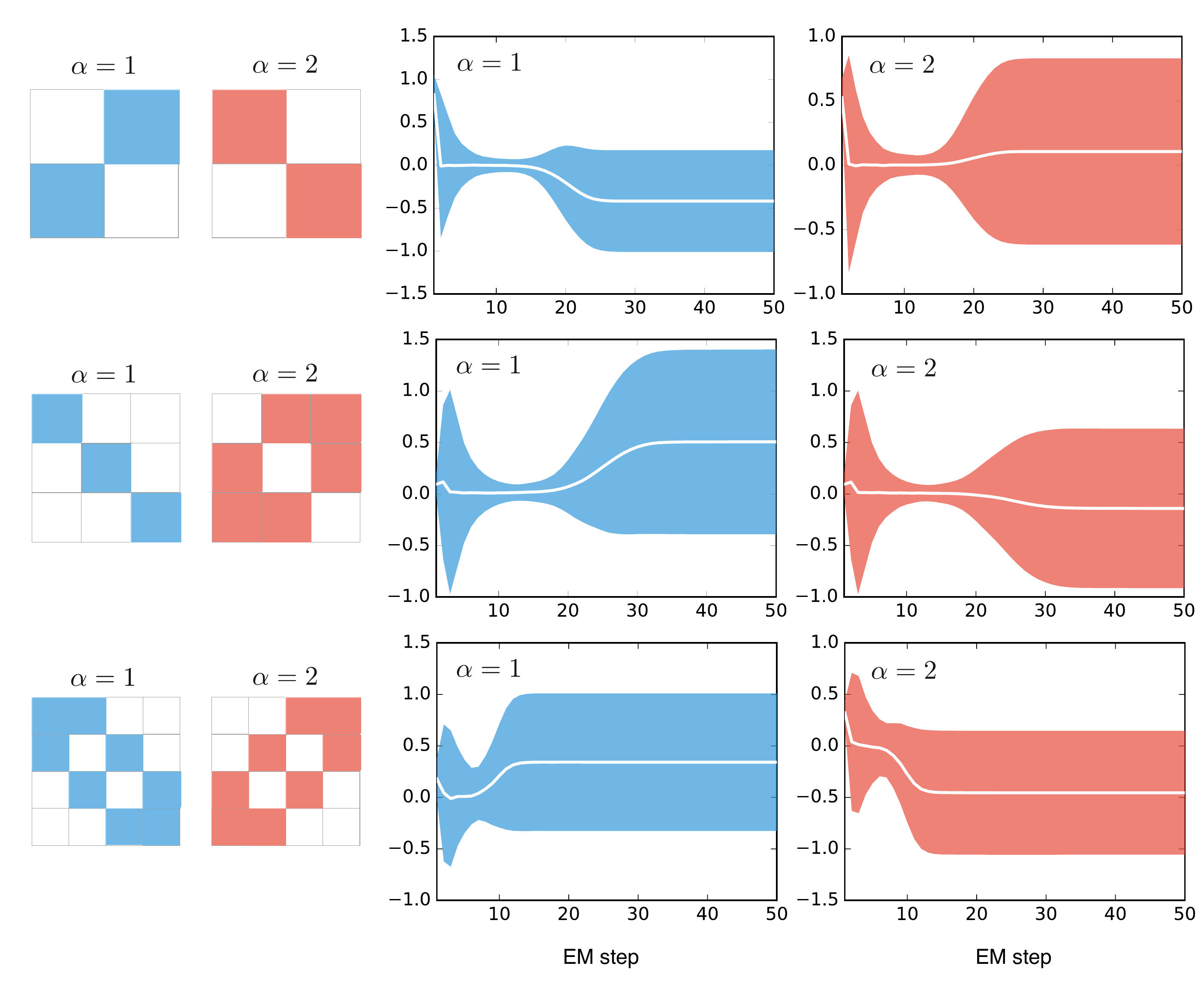}
 \end{center}
 \caption{
	(Color online) 
	Planted $W^{\alpha}$ matrices (two left panels) and evolution of the deviation from the uninformative module-assignment distribution (two right panels). 
	In each $W^{\alpha}$ matrix, the colored elements represent $W^{\alpha}_{\sigma \sigma^{\prime}} = 1$, and the white elements represent $W^{\alpha}_{\sigma \sigma^{\prime}}=0$. 
	In the right panels, the mean value $\bracket{\xi^{\alpha}_{ij}}$ is represented as a white line, and its standard deviation $\mathrm{Std}[\xi^{\alpha}_{ij}]$ is shown as a colored tie. 
	The top panels correspond to the example given in Sec.~\ref{TwoCommunityPhaseDiagram}, i.e., the simple community structure with $q=2$, and we set $(x_{1},x_{2}) = (0.1,0.6)$ as the planted model parameters. 
	The middle panels correspond to the example given in Sec.~\ref{ThreeCommunityPhaseDiagram}, i.e., the simple community structure with $q=3$, and we set $(x_{1},x_{2}) = (0.8,0.2)$ as the planted model parameters. 
	The bottom panels correspond to the example given in Sec.~\ref{FourGMGMPhaseDiagram}, i.e., the general structure of Fig.~\ref{EMPhaseDiagram-GMGM}a, and we set $(x_{1},x_{2}) = (0.8,0.1)$ as the planted model parameters. 
	}
 \label{PSIdeviations}
\end{figure*}

In Fig.~\ref{PSIdeviations}, each panel shows the evolution of $\bracket{\xi^{\alpha}_{ij}}$, i.e., the mean deviation of the module assignment distributions from the uninformative ones, and its standard deviations. 
The left and right panels represent the evolutions with respect to $\alpha=1$ and $\alpha=2$, respectively. 
To make the randomization process visible, we set $\bracket{\xi^{\alpha}_{ij}} \ne 0$ at the beginning of the algorithm, unlike our assumption in the main text. 
We can confirm that in all cases, $\bracket{\xi^{\alpha}_{ij}}$ quickly approaches zero, stays there for a moment, and eventually converges to a nontrivial value. 
These results indicate that the assumption $\bracket{\xi^{\alpha}_{ij}} = 0$ during the transient dynamics is indeed correct. 

\bibliographystyle{apsrev}
\bibliography{bib-EMdetectability}



\end{document}